\documentclass[aps,amsmath,amssymb,prl,showpacs,floatfix]{revtex4}
\usepackage{bm,natbib}
\usepackage[dvips]{graphicx}

\begin{document}

\title{Stability of Quantum Critical Points in the Presence of Competing Orders}

\author{Jian-Huang She$^{1,2}$, Jan Zaanen$^2$, Alan R. Bishop$^1$ and Alexander V. Balatsky$^1$}

\affiliation{$^1$ Theory Division, MS B 262, Los Alamos National Laboratory, Los Alamos, NM 87545, USA}

\affiliation{$^2$ Instituut-Lorentz for Theoretical Physics, Universiteit Leiden, P. O. Box 9506, 2300 R A Leiden, The Netherlands}

\begin{abstract}

We investigate the stability of Quantum Critical Points (QCPs) in the presence of two competing phases. These phases near QCPs are assumed to be either classical or quantum and assumed to repulsively interact via square-square interactions. We find that for any dynamical exponents and for any dimensionality strong enough interaction renders QCPs unstable, and drives transitions to become first order. We  propose that this instability and the onset of first-order transitions lead to spatially inhomogeneous states in practical materials near putative QCPs. Our analysis also leads us to suggest that there is a breakdown of Conformal Field Theory (CFT) scaling in the Anti de Sitter models, and in fact these models contain first-order transitions in the strong coupling limit. 
\end{abstract}

\date{\today \ [file: \jobname]}

 \maketitle

\section{Introduction}

Quantum criticality is an important concept that has dominated the landscape of modern condensed matter physics for the last decade \cite{Sachdev99}. The idea behind quantum criticality is simple and powerful. Imagine competing interactions that typically drive the transitions between different phases.  Logically one has to allow for the possibility that the relative strength of these competing interactions is tunable as a function of the external control parameters such as pressure, magnetic field, doping: we deliberately omit temperature as a control parameter since quantum phase transitions (QPTs) will occur at T=0. The simplest route to arrive at a QPT is to consider a line of finite temperature phase transition as a function of some control parameter,  such as pressure $P$, magnetic field $B$ or doping $x$. At $T=0$ this line will indicate a critical value of the control parameter. This specific value of the control parameter, where one expects a precise balance between tendency to different phases or states, is called a quantum critical point (QCP). Near this point, competing interactions nearly compensate each other. It is often asserted that it is the physics of frustration and competition, which leads to the finite temperature transition, that also controls and enables the interesting properties of materials that are brought to the $T=0$ QCP.

Much of the attention on quantum criticality has been focused on the finite temperature scaling properties \cite{Sachdev99, Continentino01, Lohneisen07}. Temperature is the only relevant scale in the quantum critical region above the QCP, bounded by the crossover line $T^*\sim |r|^{\nu z}$. The parameter $r$ measures the distance to the QCP, $\nu$ is the correlation-length exponent in $\xi\sim r^\nu$ and $z$ is the dynamical exponent in $\xi_{\tau}\sim\xi^z$. With the correlation length $\xi$ and correlation time $\xi_\tau$ much larger than any other scale of the system, power law behavior is expected for many physical observables, e. g. the specific heat, magnetic susceptibility, and most notably resistivity. Clear deviations from the Fermi liquid predictions are experimentally detected, and these phases are commonly termed non-Fermi liquids. In many systems, the anomalous finite temperature scaling properties are asserted to result from the underlying zero temperature QCPs.

In this paper, we would like to emphasize another aspect of quantum criticality, namely that it serves as a driving force for new exotic phenomena at extremely low temperatures and in extremely clean systems. One possibility is the appearance of new phases around the QCPs. It has been found in numerous experiments as one lowers temperature, seemingly inevitably in all the systems available, new phases appear near the QCP. Most commonly observed to date is the superconducting phase. The phenomenon of a superconducting dome enclosing the region near the QCP is quite general (see Fig. 1). It has been identified in many heavy fermion systems \cite{Lonzarich98, Stewart06, Gegenwart08}, plausibly also in cuprates \cite{Sachdev03}, even possibly in pnictides \cite{Zhao08, XHChen08, XHChen09, Fisher09, Canfield08, Ning09}, and probably in organic charge-transfer salts \cite{Brown00, Itoi07, Itoi0702}. Other examples include the nematic phase around the metamagnetic QCP in the bilayer ruthenate $\rm Sr_3Ru_2O_7$ \cite{Mackenzie01,Mackenzie04,Mackenzie07,Mackenzie09}, the origin of which is still under intense debate \cite{Kee05, Simons09, Kivelson09, Wu09, Sigrist10}. The emerging quantum paraelectric - ferroelectric phase diagram is also very reminiscent \cite{Millis03, Rowley09}, as is the disproportionation-superconducting phase in doped bismuth oxide superconductors \cite{Cava88, Johnson88, Pei90, Goodenough90, Allen97, Allen02}.   
\begin{figure}[ht]
\begin{center}
\includegraphics[width=7cm, clip]{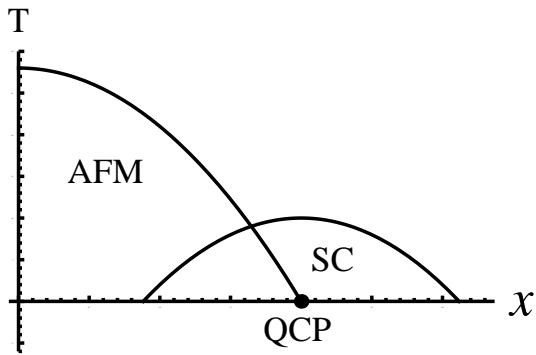}
\end{center}
\caption{Illustration of  the competing phases and superconducting dome. Here for concreteness, we consider the ordered phase to be an antiferromagnetic phase. $x$ is the tuning parameter. It can be pressure, magnetic field or doping. The superconducting temperature usually has the highest value right above the QCP.}
\label{figqcp}
\end{figure}

It has also been discovered recently that, as samples are becoming cleaner, on the approach to QCP we encounter first order transitions, and the new phases near the QCP are usually inhomogeneous and exhibit finite wavevector orderings (see \cite{Pfleiderer05, Pfleiderer09, Rosenbaum09} and references therein). For example, the heavy fermion compound $\rm CeRhIn_5$ orders antiferromagnetically at low temperature and ambient pressure. As pressure increases, the Neel temperature decreases and at some pressure the antiferromagnetic phase is replaced by a superconducting phase through a first-order phase transition. There are also evidences for a competitive coexistence of the two phases within the antiferromagnetic phase, as in some organic charge-transfer superconductor precursor antiferromagnetic phases. Such coexistence was also observed in $\rm Rh$-doped $\rm CeIrIn_5$. The heavy fermion superconductor $\rm CeCoIn_5$ has the unusual property that when a magnetic field is applied to suppress superconductivity, the superconducting phase transition becomes first-order below $T_0\simeq 0.7 K$. For the superconducting ferromagnet $\rm UGe_2$, where superconductivity exists within the ferromagnetic state, the two magnetic transitions (ferromagnetic to paramagnetic and large-moment ferromagnetic to small-moment ferromagnetic) are both first order \cite{Flouquet00, Huxley01, Huxley02}. Other examples of continuous phase transitions turning first-order at low temperatures include $\rm CeRh_2Si_2$ \cite{Roman97, Roman96}, $\rm CeIn_3$ \cite{Onuki08}, $\rm URhGe$ \cite{Levy05}, $\rm ZrZn_2$ \cite{Uhlarz04} and $\rm MnSi$ \cite{Pfleiderer07}. The prevailing point of view seems to be that  this happens only in a few cases and these are considered exceptions. Yet we are facing a rapidly growing list of these "exceptions", and we take the view here that they rather represent a general property of QCPs. 

The point is that, on approach to the QCP, an interaction that was deemed irrelevant initially, takes over and dominates. For example it has been proposed recently that the superconducting instability, which is marginal in the usual Fermi liquids, becomes relevant near the QCP and leads to a high transition temperature \cite{She09}. Actually these instabilities are numerous and can vary, depending on the system at hand. However there seems to be a unifying theme of those instabilities. We suggest that  QCPs are unstable precisely for the reasons we are interested in these points: extreme softness and extreme susceptibility of the system in the vicinity of QCPs. We regard the recently discovered first order transitions as indicators of a more fundamental and thus powerful phsyics. We are often prevented from reaching quantum criticality, and often the destruction is relatively trivial and certainly not as appealing and elegant as quantum criticality. We can draw an analogy from gravitational physics, where the naked singularities are believed to be prevented from happening due to many kinds of relevant instabilities. This is generally known as the "cosmic censorship conjecture" \cite{Penrose69}. The recently proposed AdS/CFT correspondence \cite{Maldacena97, Witten98, Gubser98}, which maps a non-gravitational field theory to a higher dimensional gravitational theory, adds more to this story. Here researchers have begun to realize that the Reissner-Nordstrom black holes in AdS space, which should have a macroscopic entropy at zero temperature, are unstable to the spontaneous creation of particle-antiparticle pairs, and tend to collapse to a state with lower entropy \cite{Hartnoll08, Zaanen0902, Sachdev10}.

There have appeared in the literature scattered examples of first-order quantum phase transitions at the supposed-to-be continuous QCPs \cite{Belitz05,Continentino01,Continentino04,Continentino0402,Continentino0403,Continentino05,Qi09,Millis10}, however it appears that the universality of this phenomenon is not widely appreciated. This universality is the main motivation for our paper. We will systematically study the different possibilities for converting a continuous QPT to first order.

The first striking example how fluctuations of one of the order parameters can qualitatively change the nature of the transition comes from the Coleman-Weinberg model \cite{Coleman73}, where they showed how gauge fluctuations of the charged field introduce a first order transition. In this work it was shown that in dimension $d= 3$, for any weak coupling strength, one develops a logarithmic singularity, and therefore the effective field theory has a first-order phase transition. Subsequently, this result was extended to include classical gauge field fluctuations by Halperin, Lubensky and Ma \cite{Halperin74}, where a cubic correction to the free energy was found. Nontrivial gradient terms can also induce an inhomogeneous phase and/or glassy behavior \cite{Sasha09}.

A prototypical example for the competing phases and superconducting dome is shown schematically in Fig.1.  Below, we apply the renormalization group (RG) and scaling analysis to infer the stability if the QCP as a result of competition. We find in our analysis that the QCP is indeed unstable towards a first order transition as a result of competition. Obviously details of the collapse of a QCP and the resulting phase diagram depend on details of the nature of the fluctuating field and details of the interactions. We find that the most relevant parameters that enter into criterion for stability of a QCP are the strength of interactions between competing phases: we take this interaction to be repulsive between squares of the competing order parameters. When the two order parameters break different symmetries, the coupling will be between the squares of them.  Another important factor that controls the phase diagram is the dynamical exponents $z$ of the fields. The nature of the competition also depends on the classical or quantum character of the fields. Here by classical we do not necessarily mean a finite temperature phase transition, but rather that the typical energy scale is above the ultraviolet cutoff, and the finite frequency modes of the order parameters can be ignored, so that a simple description in terms of free energy is enough to capture the physics. We analyzed three possibilities for the competing orders:

i) \emph{classical} + \emph{classical}. Here we found that interactions generally reduce the region of coexistence, and when interaction strength exceeds some critical value, the second-order phase transitions become first order.

ii) \emph{classical} + \emph{quantum}. Here the quantum field is integrated out, giving rise to a correction to the effective potential of the classical order parameter. For a massive fluctuating field with $d+z\leqslant 6$, or a massless one with $d+z\leqslant 4$, the second-order quantum phase transition becomes first order.

iii) \emph{quantum} +  \emph{quantum}. Here RG analysis was employed, and we found that in the high dimensional parameter space, there are generally regions with runaway flow, indicating a first-order quantum phase transition.

It has been proposed recently that alternative route to the breakdown of quantum criticality is through the basic collapse of  Landau-Wilson paradigm of conventional order parameters and formation of the deconfined quantum critical phases (\cite{Senthil04, Senthil0402}). This is a possibility that has been discussed for specific models and requires a different approach than the one taken here. We are not addressing this possibility.

The plan of the paper is as follows.  In section II, we consider coupling two classical order parameter fields together. Both fields are characterized by their free energies and Landau mean field theory will be used. In section III, we consider coupling a classical order parameter to a quantum mechanical one, which can have different dynamical exponents. The classical field is described by its free energy and the quantum field by its action; the latter is integrated out to produce a correction to the effective potential for the former. In section IV, we consider coupling two quantum mechanical fields together. With both fields described by their actions, we use RG equations to examine the stability conditions. In particular, we study in detail the case where the two coupled order parameters have different dynamical exponents, which, to our knowledge, has not been considered previously. In the conclusion section, we summarize our findings.

\section{Two competing classical fields}
 We consider in this section two competing classical fields. Examples are the superconducting order and antiferromagnetic order in $\rm CeRhIn_5$ and  $\rm Rh$-doped $\rm CeIrIn_5$, and the superconducting order and ferromagnetic order near the large-moment to small-moment transition in $\rm UGe_2$. We will follow the standard textbook approach, and this case is presented as a template for the more complex problems studied later on.

 We first study the problem at zero temperature. For simplicity, both of them are assumed to be real scalars. The free energy of the system consists of three parts, the two free parts $F_{\psi}, F_M$ and the interacting part $F_{\rm int}$:
\begin{equation}
 \begin{split}
F=&F_{\psi}+F_M+F_{\rm int};
\\F_{\psi}=&\frac{\rho}{2}(\nabla \psi)^2-\alpha \psi^2+\frac{\beta}{2}\psi^4;
\\ F_M=&\frac{\rho_M}{2}(\nabla M)^2-\alpha_M M^2+\frac{\beta_M}{2}M^4;
\\F_{\rm int}=&\gamma \psi^2M^2 .
\end{split}
\label{freeenergy}
\end{equation}
Here, by changing $\alpha, \alpha_M$, the system is tuned through the phase transition points.
When the two fields are decoupled, with $\gamma=0$, there will be two separated second-order phase transitions. Assume the corresponding values of the tuning parameter $x$ at these two transition points are $x_1$ and $x_2$, we can parameterize $\alpha, \alpha_M$ as $\alpha=a(x-x_1)$ and $\alpha_M=a_M(x_2-x)$, where $a,a_M$ are constants.

\begin{figure}[ht]
\begin{center}
\includegraphics[width=5.2cm, clip]{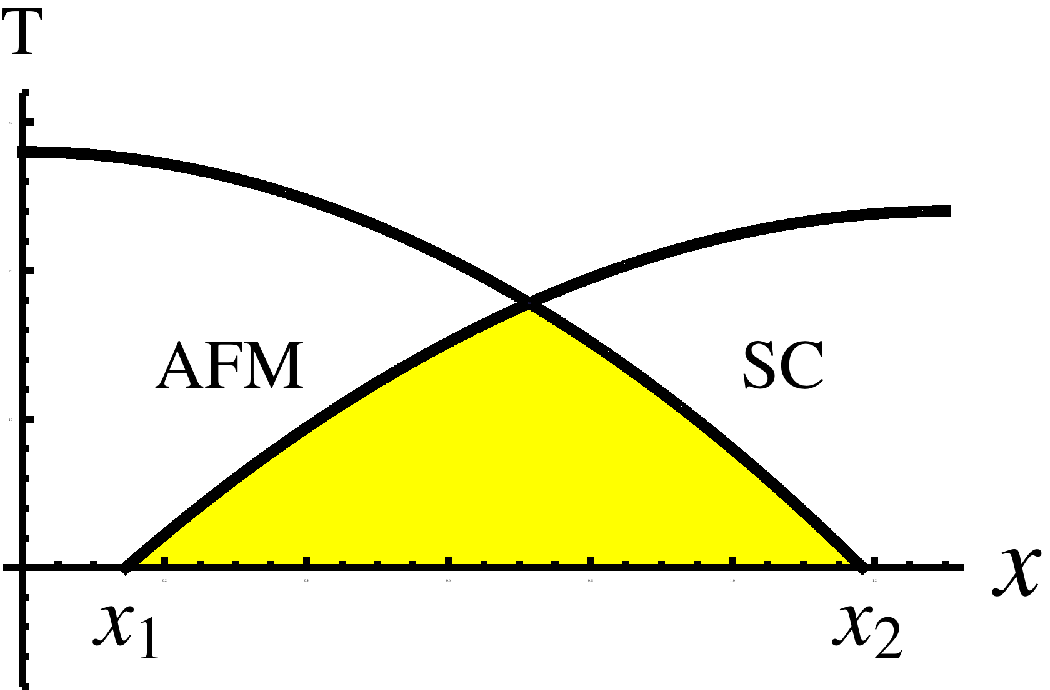}
\includegraphics[width=5.2cm, clip]{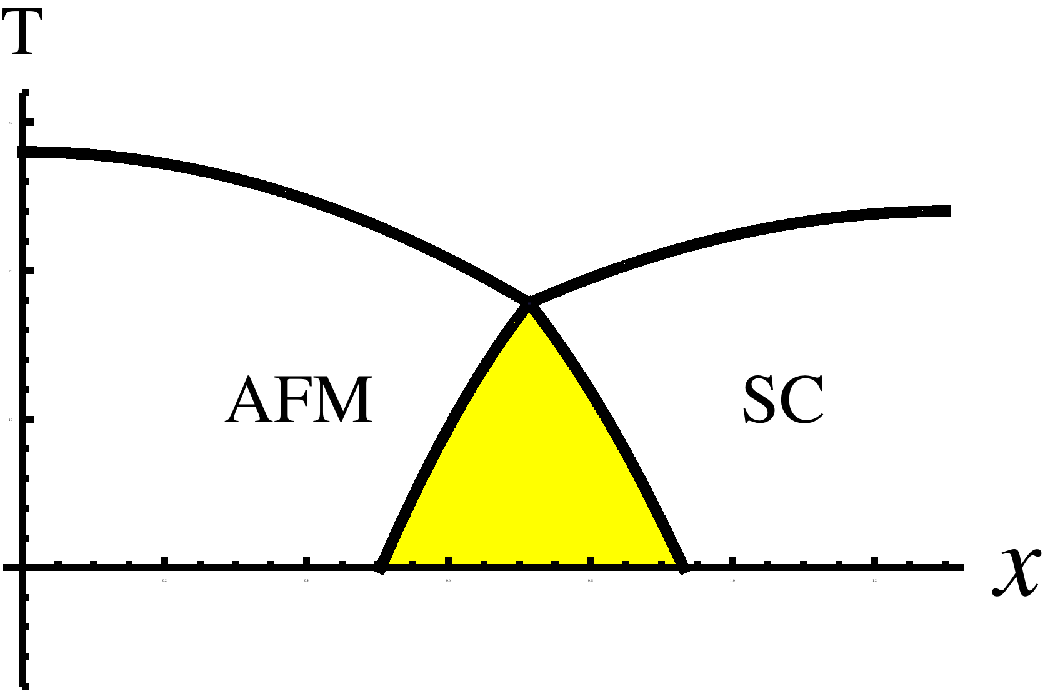}
\includegraphics[width=5.2cm, clip]{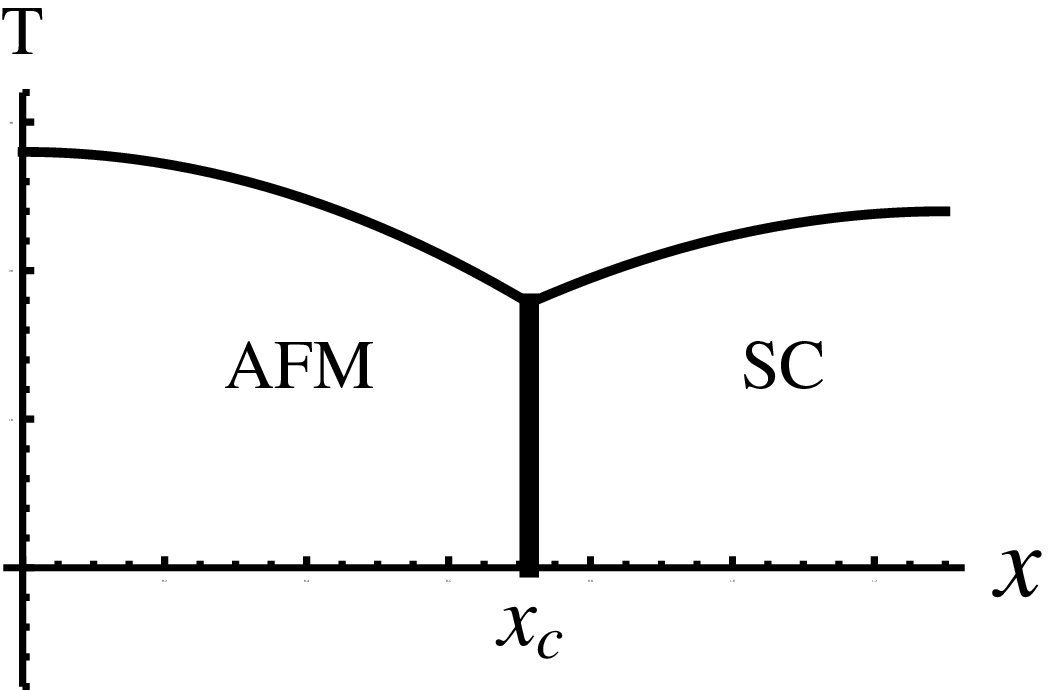}
\end{center}
\caption{(Color online) Illustration of the mean field phase diagram for two competing orders. Here for concreteness we consider antiferromagnetic and superconducting orders.  The two orders coexist in the yellow region, whose area shrinks as the coupling increases from left to right.  The left figure has $\gamma=0$, the central one has $0<\gamma<\sqrt{\beta\beta_M}$, and the right one has $\gamma>\sqrt{\beta\beta_M}$. When $\gamma$ exceeds the critical value $\sqrt{\beta\beta_M}$, the two second-order phase transition lines merge and become first order (the thick vertical line).}
\label{fig1}
\end{figure}

We would like to know the ground state of the system. Following the standard procedure, we first find the homogeneous field configurations satisfying $\frac{\partial F}{\partial \psi}=\frac{\partial F}{\partial M}=0$, and then compare the corresponding free energy. It is easy to see that the above equations have four solutions, with $(|\psi|, |M|)=(0,0), (0,\sqrt{\alpha_M/\beta_M}), (\sqrt{\alpha/\beta},0),(\psi_*,M_*)$, where \begin{equation}
 \begin{split}
\alpha \psi_*^2=&\frac{\gamma'-\beta'_M}{{\gamma'}^2-\beta'\beta'_M},
\\ \alpha_M M_*^2=&\frac{\gamma'-\beta'}{{\gamma'}^2-\beta'\beta'_M},
\end{split}
\label{psiM}
\end{equation}
and the rescaled parameters are $\gamma'=\gamma/\alpha\alpha_M, \beta'=\beta/\alpha^2,\beta'_M=\beta_M/\alpha_M^2$. When $\gamma=0$, the fourth solution reduces to $(\psi_*, M_*)=(\sqrt{\alpha/\beta},\sqrt{\alpha_M/\beta_M})$, with the two orders coexisting but decoupled. We are interested in the case where the two orders are competing, thus a relatively large positive $\gamma$.

For $x_1<x<x_2$, we have $\alpha>0, \alpha_M>0$. The necessary condition for the existence of the fourth solution is $\gamma'>\beta', \beta'_M, \sqrt{\beta'\beta'_M}$ or $\gamma'<\beta', \beta'_M, \sqrt{\beta'\beta'_M}$. In this case, the configuration $(0,0)$ has the highest free energy $F[0,0]=0$.
For the configuration $(\psi_*,M_*)$ with coexisting orders to have lower free energy than the two configurations with single order, one needs to have $\gamma'<\sqrt{\beta'\beta'_M}$, which reflects the simple fact that when the competition between the two orders is too large, their coexistence is not favored. Thus the condition for the configuration $(\psi_*,M_*)$ to be the ground state of the system is $\gamma'<\beta'$ and $\gamma'<\beta'_M$. If $\gamma'>{\rm min}\{\beta',\beta'_M\}$, one of the fields has to vanish. 

Next we observe that, for $x$ near $x_1$, $\beta'_M$ remains finite, $\alpha \sim (x - x_1)$, and $\gamma'$ diverges as $1/(x-x_1)$, while $\beta'$ diverges as $1/(x-x_1)^2$. So the lowest energy configuration is $\psi=0$, $|M|=\sqrt{\alpha_M/\beta_M}$. Similarly, near $x_2$, the ground state is $(\sqrt{\alpha/\beta},0)$. The region with coexisting orders shrinks to
\begin{equation}
\frac{\gamma a_Mx_2+\beta_M ax_1}{\gamma a_M+\beta_M a}<x<\frac{\gamma ax_1+\beta a_Mx_2}{\gamma a+\beta a_M}.
\end{equation}
For $\gamma<\sqrt{\beta\beta_M}$, this region has finite width. In this region, $(0,0)$ is the global maximum of the free energy, $(0,\sqrt{\alpha_M/\beta_M}), (\sqrt{\alpha/\beta},0)$ are saddle points, and $(\psi_*, M_*)$ is the global minimum. The phase with coexisting order is sandwiched between the two singly ordered phases, and the two phase transitions are both second-order. The shift in spin-density wave ordering and Ising-nematic ordering due to a nearby competing superconducting order has been studied recently by Moon and Sachdev \cite{Moon09, Moon10}, where they found that the fermionic degrees of freedom can play important roles.

For $\gamma>\sqrt{\beta\beta_M}$, this intermediate region with coexisting orders vanishes, and the two singly ordered phases are separated by a first-order quantum phase transition. The location of the phase transition point is determined by equating the two free energies at this point,
\begin{equation}
F\left[\sqrt{\frac{\alpha(x_c)}{\beta}},0\right]=F\left[0,\sqrt{\frac{\alpha_M(x_c)}{\beta_M}}\right],
\end{equation}
which gives $x_c=(x_2+Ax_1)/(1+A)$, with $A=(a/a_M)\sqrt{\beta_M/\beta}$. The slope of the free energy changes discontinuously across the phase transition point, with a jump
\begin{equation}
\delta F^{(1)}\equiv\left\vert\left(\frac{dF}{dx}\right)_{x_c^+}-\left(\frac{dF}{dx}\right)_{x_c^-}\right\vert=\frac{aa_M}{\sqrt{\beta\beta_M}}(x_2-x_1).
\end{equation}

The size of a first-order thermal phase transition can be characterized by the ratio of latent heat to the jump in specific heat in a reference second-order phase transition \cite{Halperin74}. A similar quantity can be defined for a quantum phase transition, where the role of temperature is now played by the tuning parameter $x$. We choose as our reference point $\gamma=0$, where the two order parameters are decoupled. For $x<x_1$, one has $d^2F/dx^2=-a_M^2/\beta_M$; for $x>x_2$, one has $d^2F/dx^2=-a^2/\beta$; and $d^2F/dx^2=-a_M^2/\beta_M-a^2/\beta$ for $x_1<x<x_2$. We take the average of the absolute value of the two jumps to obtain
\begin{equation}
\delta F^{(2)}=\frac{1}{2}(a_M^2/\beta_M+a^2/\beta).
\end{equation}
So the size of this first-order quantum phase transition is 
\begin{equation}
\delta x=\frac{\delta F^{(1)}}{\delta F^{(2)}}=\frac{2\sqrt{\tilde{\beta}\tilde{\beta}_M}}{\tilde{\beta}+\tilde{\beta}_M}(x_2-x_1),
\end{equation}
with $\tilde{\beta}=\beta/a^2$ and $\tilde{\beta}_M=\beta_M/a_M^2$. It is of order $x_2-x_1$, when $\tilde{\beta}$ and $\tilde{\beta}_M$ are not hugely different.

The above consideration can be generalized to finite temperature, by including the temperature dependence of all the parameters. Specially, there exists some temperature $T^*$, where $x_1(T^*)=x_2(T^*)$. In this way we obtain phase diagrams similar to those observed in experiments (see Fig. 1). 

\section{Effects of quantum fluctuations}
In this section, we consider coupling an order parameter $\psi$ to another field $\phi$, which is fluctuating quantum mechanically. The original field $\psi$ is still treated classically, meaning any finite frequency modes are ignored. For the quantum fields, in the spirit of Hertz-Millis-Moriya \cite{Hertz76, Millis93, Moriya85}, we assume that  the fermionic degrees of freedom can be integrated out, and we will only deal with the bosonic order parameters. This model may, for example, explain the first-order ferromagnetic to paramagnetic transition in $\rm UGe_2$, where the quantum fluctuations of the superconducting order parameter are coupled with the  ferromagnetic order parameter, which can be regarded as classical near the superconducting transition point.

 We will integrate out the quantum field to obtain the effective free energy of a classical field. The partition function has the form
\begin{equation}
Z[\psi(\mathbf r)]=\int {\cal D}\phi({\mathbf r},\tau) \exp\left({-\frac{{\cal F}_{\psi}}{T}-S_{\phi}-S_{\psi\phi}}\right).
\end{equation}
The free energy is of the same form as in the previous section with ${\cal F}_{\psi}=\int d^d{\mathbf r}F_{\psi}$. Thus, in the absence of coupling to other fields, the system goes through a second-order quantum phase transition as one tunes the control parameter $x$ across its critical value. We consider a simple coupling
\begin{equation}
S_{\psi\phi}=g\int d^d{\mathbf r}d\tau \psi^2\phi^2.
\end{equation}
The action of the $\phi$ field depends on its dynamical exponent $z$. We notice that such classical + quantum formalism has been used to investigate the competing orders in cuprates in \cite{Sachdev02}.

 The saddle point equation for $\psi$ reads
\begin{equation}
\frac{\delta\ln Z[\psi(\mathbf r)]}{\delta\psi(\mathbf r)}=0,
\end{equation}
which gives
\begin{equation}
\left[-\alpha+\beta\psi^2(\mathbf r)-\frac{\rho}{2}\nabla^2+g\langle\phi^2(\mathbf r)\rangle\right]\psi(\mathbf r)=0.
\end{equation}
Here we have defined the expectation value,
\begin{equation}
\langle\phi^2(\mathbf r)\rangle=\frac{1}{\beta}\int {\cal D}\phi({\mathbf r'},\tau') \int_0^{\beta}d\tau\phi^2({\mathbf r},\tau) \exp\left({-S_{\phi}-S_{\psi\phi}}\right).
\end{equation}
It can also be written in terms of the different frequency modes,
\begin{equation}
\langle\phi^2(\mathbf r)\rangle=T\sum_{\omega_n}\langle\phi(\mathbf r,\omega_n)\phi(\mathbf r,-\omega_n)\rangle=T\sum_{\omega_n}\int {\cal D}\phi({\mathbf r'},\nu_s)\phi(\mathbf r,\omega_n)\phi(\mathbf r,-\omega_n) \exp\left({-S_{\phi}-S_{\psi\phi}}\right).
\end{equation}
The quadratic term in $S_{\phi}$ is of the form 
\begin{equation}
S_{\phi}^{(2)}=\sum_{\nu_s}\int d^d{\mathbf r}' \int d^d{\mathbf r}'' \phi({\mathbf r'},\nu_s)\chi_0^{-1}({\mathbf r'},{\mathbf r''},\nu_s)\phi(\mathbf r'',-\nu_s), 
\end{equation}
or more conveniently, in terms of momentum and frequency,
\begin{equation}
S_{\phi}^{(2)}=\sum_{\nu_s}\int \frac{d^d\mathbf k}{(2\pi)^d} \phi({\mathbf k},\nu_s)\chi_0^{-1}({\mathbf k},\nu_s)\phi(-\mathbf k,-\nu_s).
\end{equation}
So in the presence of translational symmetry, we find
\begin{equation}
\langle\phi^2\rangle=T\sum_{\omega_n}\int  \frac{d^d\mathbf k}{(2\pi)^d} \frac{1}{\chi_0^{-1}({\mathbf k},\omega_n)+g\psi^2}.
\end{equation}
This leads to the 1-loop correction to the effective potential for $\psi$, determined by
\begin{equation}
\frac{\delta V_{\rm eff}^{(1)}[\psi]}{\delta\psi}=2g\langle\phi^2\rangle\psi.
\end{equation}

So far we have been general in this analysis. Further analysis requires us to make more specific assumptions about the dimensionality and dynamical exponents.

\subsection{Fluctuations with $d=3, z=1$}
When the $\phi$ field has dynamical exponent $z=1$, its propagator is of the form
\begin{equation}
\chi_0({\mathbf k},\omega_n)=\frac{1}{\omega_n^2+k^2+\xi^{-2}}.
\end{equation}
A special case is a gauge boson, which has zero bare mass, and thus $\xi\to\infty$. This problem has been studied in detail by Halperin, Lubensky and Ma \cite{Halperin74} for a classical phase transition (see also \cite{Lubensky78}), and by Coleman and Weinberg \cite{Coleman73} for relativistic quantum field theory. Other examples are critical fluctuations associated with spin-density wave transitions and superconducting transitions in clean systems.
We also note that Continentino and collaborators have used the method of effective potential to investigate some special examples of the fluctuation-induced first order quantum phase transition \cite{Continentino01,Continentino04,Continentino0402,Continentino0403,Continentino05}.
\begin{figure}[ht]
\begin{center}
\includegraphics[width=6cm, clip]{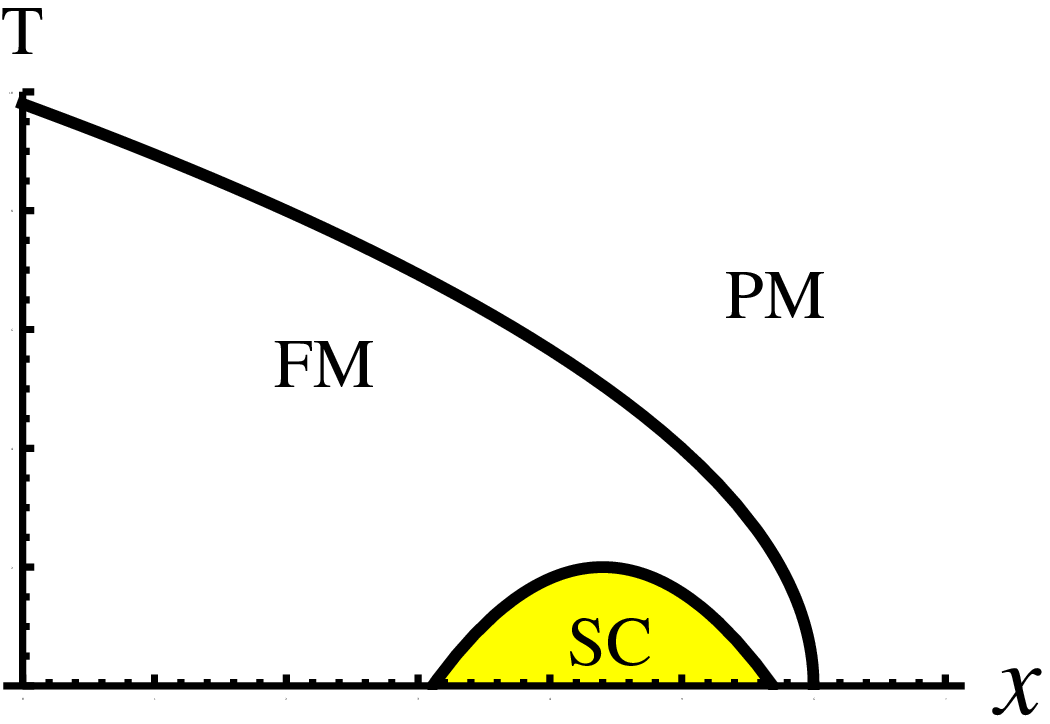}
\includegraphics[width=6cm, clip]{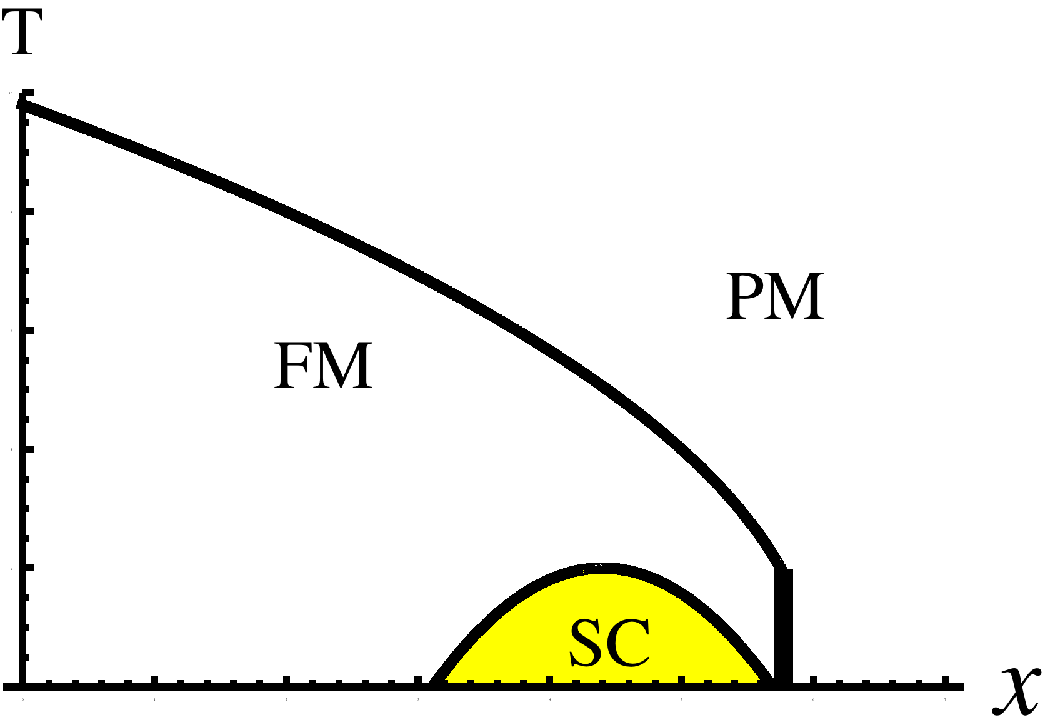}
\end{center}
\caption{(Color online) Schematic illustration of the fluctuation-induced first-order phase transition. Here, for concreteness, we consider ferromagnetic and superconducting orders. The ferromagnetic order is regarded as classical, while the superconducting one as quantum mechanical. At low temperatures, the second-order ferromagnetic to paramagnetic phase transition becomes first order (the thick vertical line), due to fluctuations of the superconducting order parameter.}
\label{fig2}
\end{figure}

Let us consider $T=0$, for which the summation $T\sum_{\omega_n}$ can be replaced by the integral $\int d\omega/(2\pi)$. We then get for the one-loop correction to the effective potential
\begin{equation}
\frac{\delta V_{\rm eff}^{(1)}[\psi]}{\delta\psi}=2g\psi\int \frac{d\omega}{2\pi}\int \frac{d^d{\mathbf k}}{(2\pi)^d}\frac{1}{\omega^2+k^2+\xi^{-2}+g\psi^2}.
\end{equation}
Carrying out the frequency integral, we obtain for $d=3$,
\begin{equation}
\frac{\delta V_{\rm eff}^{(1)}[\psi]}{\delta\psi}=\frac{g\psi}{2\pi^2}\int_0^{\Lambda}dk\frac{k^2}{\sqrt{k^2+\xi^{-2}+g\psi^2}},
\end{equation}
where an ultraviolet cutoff is imposed. Integrating out momentum gives
\begin{equation}
\frac{\delta V_{\rm eff}^{(1)}[\psi]}{\delta\psi}=\frac{g\psi}{4\pi^2}\left[\Lambda\sqrt{\Lambda^2+\xi^{-2}+g\psi^2}-(\xi^{-2}+g\psi^2)\ln\left(\frac{\Lambda+\sqrt{\Lambda^2+\xi^{-2}+g\psi^2}}{\sqrt{\xi^{-2}+g\psi^2}}\right)\right],
\end{equation}
which can be simplified as
\begin{equation}
\frac{\delta V_{\rm eff}^{(1)}[\psi]}{\delta\psi}=\frac{g\psi}{4\pi^2}\left[\Lambda^2+\frac{1}{2}(\xi^{-2}+g\psi^2)-(\xi^{-2}+g\psi^2)\ln\left(\frac{2\Lambda}{\sqrt{\xi^{-2}+g\psi^2}}\right)\right].
\end{equation}
Combined with the bare part,
\begin{equation}
V_{\rm eff}^{(0)}(\psi)=-\alpha\psi^2+\frac{1}{2}\beta\psi^4,
\end{equation}
we get the effective potential to one-loop order,
\begin{equation}
V_{\rm eff}(\psi)=-{\hat \alpha}\psi^2+\frac{1}{2}{\hat \beta}\psi^4-\frac{1}{16\pi^2}(\xi^{-2}+g\psi^2)^2\ln\left(\frac{2\Lambda}{\sqrt{\xi^{-2}+g\psi^2}}\right),
\end{equation}
with the quadratic and quartic terms renormalized by ${\hat \alpha}=\alpha-g(4\Lambda^2+\xi^{-2})/(32\pi^2)$ and ${\hat \beta}=\beta+3g/(32\pi^2)$. When $\phi$ field is critical with $\xi\to\infty$, the third term is of the well-known Coleman-Weinberg form $\psi^4\ln(2\Lambda/\sqrt{g\psi^2})$, which drives the second-order quantum phase transition to first order. 

For $\xi$ large but finite, we can expand the third term as a power series in $\xi^{-2}/(g\psi^2)$, and the
effective potential is of the  form 
\begin{equation}
V_{\rm eff}(\psi)=-{\bar \alpha}\psi^2+\frac{1}{2}{\bar \beta}\psi^4-\frac{1}{16\pi^2}(2\xi^{-2}g\psi^2+g^2\psi^4)\ln\frac{2\Lambda}{\sqrt{g\psi^2}}.
\end{equation}
In addition to the Coleman-Weinberg term, there is another term of the form $\psi^2\ln\psi$, and again we have also a first-order phase transition.

To study the generic case where the $\phi$ field is massive, we rescale the $\psi$ field and cutoff, defining 
\begin{equation}
u^2\equiv \frac{g\psi^2}{\xi^{-2}} ,~~~~  {\tilde \Lambda}\equiv\frac{2\Lambda}{\xi^{-1}}.
\end{equation}
The rescaled effective potential takes the form
\begin{equation}
{\tilde V}_{\rm eff}(u)=-{\tilde A}u^2+\frac{1}{2}{\tilde B}u^4-(1+u^2)^2\ln\left(\frac{\tilde \Lambda}{\sqrt{1+u^2}}\right),
\end{equation}
which can be further simplified as
\begin{equation}
{\hat V}_{\rm eff}(u)=-Au^2+\frac{1}{2} Bu^4+(1+u^2)^2\ln(1+u^2).
\end{equation}
The above potential is plotted in Fig. 4.
We notice that with large enough cutoff $\Lambda$, one generally has $B={\tilde B}-\ln{\tilde \Lambda}$ large and negative. For $A<1$, $u=0$ is a local minimum. There are also another two local minima with $u^2\equiv y$ a positive solution of equation
\begin{equation}
2(1+y)\ln(1+y)+(1+B)y+1-A=0.
\end{equation}
So we generally have a first-order quantum phase transition in this case (see Fig. 3 for a schematic picture).

 \begin{figure}
\begin{center}
\includegraphics[width=0.46\linewidth]{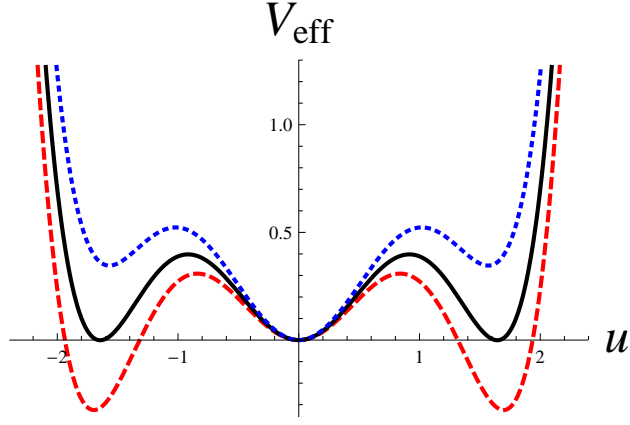}
\end{center}
\caption{(Color online) The effective potential as a function of the rescaled field $u$ for various parameters in the case $d=3, z=1$. Here $V_{\rm eff}(u)=-Au^2+\frac{1}{2} Bu^4+(1+u^2)^2\ln(1+u^2)$, with $B=-5$, and $A=-0.25, -0.116, 0$ from top to bottom.}
\end{figure}

\subsection{Fluctuations with $d=3, z=2$}
With dynamical exponent $z=2$, the propagator of $\phi$ field  is 
\begin{equation}
\chi_0({\mathbf k},\omega_n)=\frac{1}{|\omega_n|\tau_0+k^2+\xi^{-2}}.
\end{equation}
Examples are charge-density-wave and antiferromagnetic fluctuations. In the presence of dissipation, superconducting transitions also have dynamical exponent $z=2$.
 \begin{figure}
\begin{center}
\includegraphics[width=0.38\linewidth]{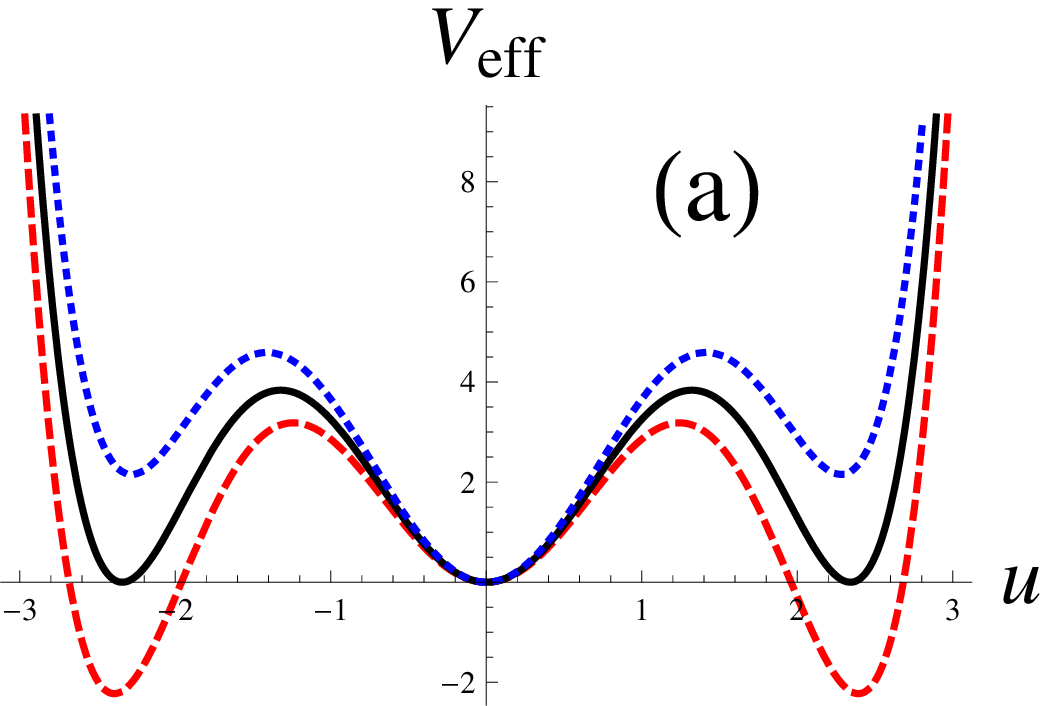}
\includegraphics[width=0.38\linewidth]{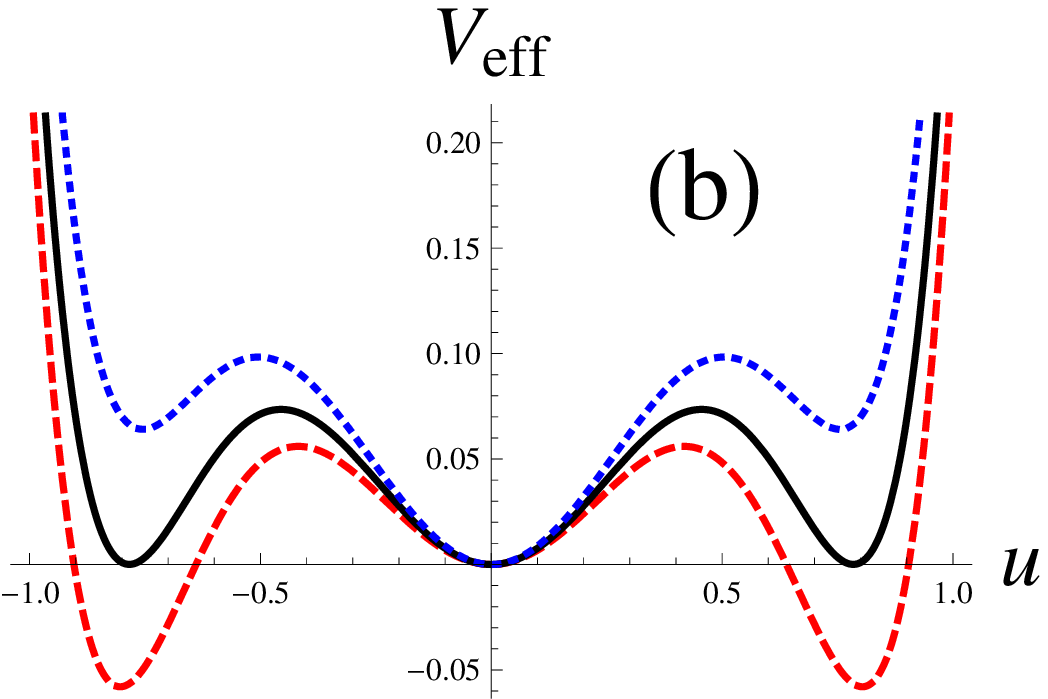}
\includegraphics[width=0.38\linewidth]{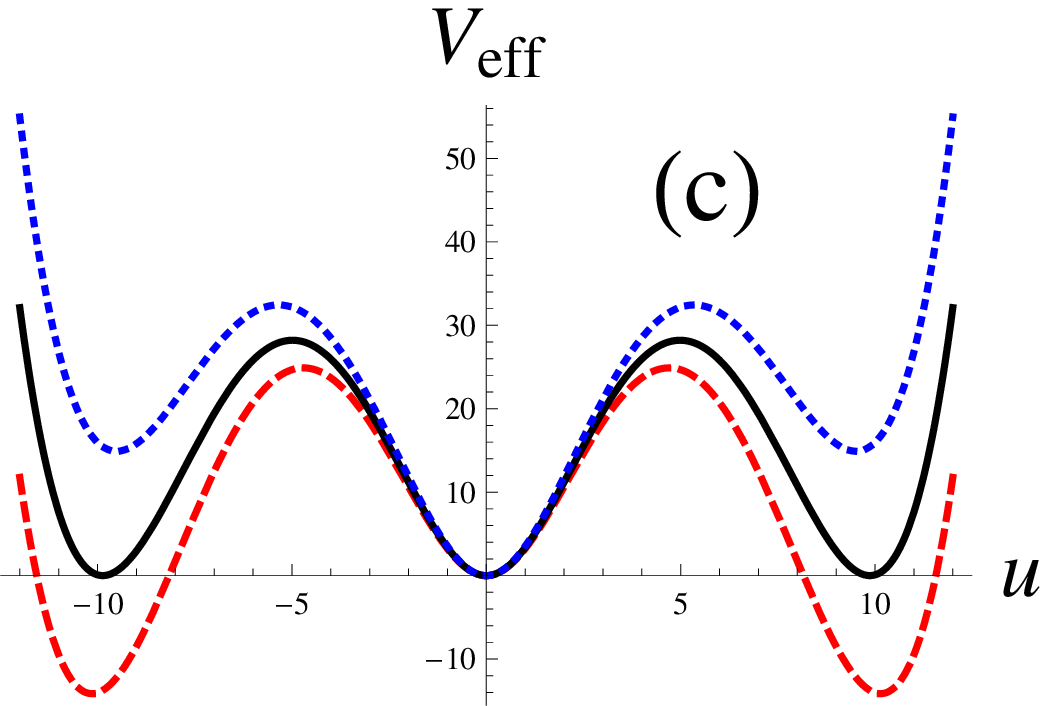}
\includegraphics[width=0.38\linewidth]{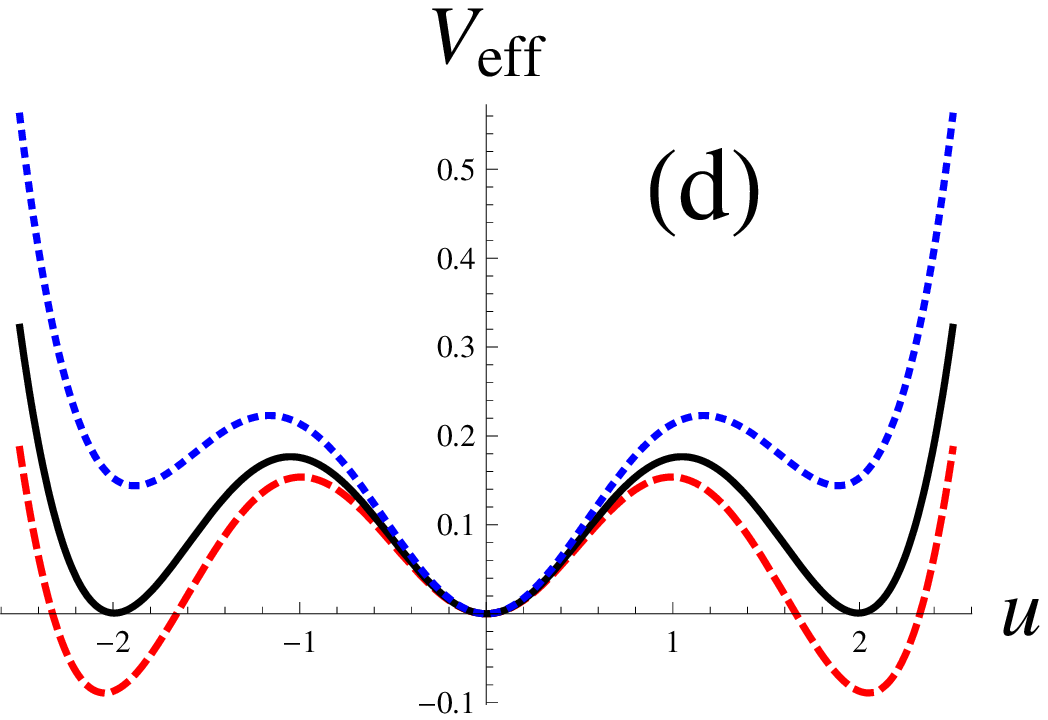}
\end{center}
\caption{(Color online) The effective potential as a function of the rescaled field $u$ for (a) $d=3, z=2$, where we have plotted $V_{\rm eff}(u)=-Au^2+\frac{1}{2} Bu^4+(1+u^2)^{5/2}-1$, with $B=-8$, and $A=-3, -2.597, -2.2$ from top to bottom; (b) $d=3, z=3$, where $V_{\rm eff}(u)=-Au^2+\frac{1}{2} Bu^4+(1+u^2)^3\ln(1+u^2)$, with $B=-10$, and $A=0.1, 0.208, 0.3$ from top to bottom;   (c) $d=1, z=2$, where we have plotted $V_{\rm eff}(u)=-Au^2+\frac{1}{2} Bu^4-(1+u^2)^{3/2}+1$, with $B=0.1$, and $A=-5.3, -5.1413, -5$ from top to bottom; (d) $d=1, z=1$, where we have plotted $V_{\rm eff}(u)=-Au^2+\frac{1}{2} Bu^4-(1+u^2)\ln(1+u^2)$, with $B=0.3$, and $A=-1.45, -1.412, -1.39$ from top to bottom. All these plots are of similar shape. However, we notice that the scales are quite different.}
\end{figure}
So the one-loop correction to the effective potential at zero temperature becomes  
\begin{equation}
\frac{\delta V_{\rm eff}^{(1)}[\psi]}{\delta\psi}=2g\psi\int \frac{d\omega}{2\pi}\int \frac{d^d{\mathbf k}}{(2\pi)^d}\frac{1}{|\omega|\tau_0+k^2+\xi^{-2}+g\psi^2}.
\end{equation}
The momentum integral is cutoff at $|\mathbf k|=\Lambda$, and correspondingly the frequency integral is cutoff at $|\omega|\tau_0=\Lambda^2$. First, we integrate out frequency to obtain
\begin{equation}
\frac{\delta V_{\rm eff}^{(1)}[\psi]}{\delta\psi}=\frac{g\psi}{\pi^3\tau_0}\int_0^{\Lambda}dkk^2\ln\left(1+\frac{\Lambda^2}{k^2+\xi^{-2}+g\psi^2}\right),
\end{equation}
and then integrate out momentum, with the final result
\begin{equation}
 \begin{split}
\frac{\delta V_{\rm eff}^{(1)}[\psi]}{\delta\psi}=\frac{g\psi}{3\pi^3\tau_0}\left[ \Lambda^3\ln\left(\frac{\xi^{-2}+g\psi^2+2\Lambda^2}{\xi^{-2}+g\psi^2+\Lambda^2}\right) +2\Lambda^3+2(\xi^{-2}+g\psi^2)^{3/2} \arctan\frac{\Lambda}{\sqrt{\xi^{-2}+g\psi^2}}\right. \\\left.-2(\xi^{-2}+g\psi^2+\Lambda^2)^{3/2} \arctan\frac{\Lambda}{\sqrt{\xi^{-2}+g\psi^2+\Lambda^2}}\right].
 \end{split}
\end{equation}
Up to order $\Lambda^0$, this is
\begin{equation}
\frac{\delta V_{\rm eff}^{(1)}[\psi]}{\delta\psi}=\frac{g\psi}{3\pi^3\tau_0}\left[ \Lambda^3\left(2+\ln 2-\frac{\pi}{2}\right)+\frac{3\pi}{4}\Lambda(\xi^{-2}+g\psi^2) +\pi(\xi^{-2}+g\psi^2)^{3/2} \right].
\end{equation}
The first two terms just renormalize the bare $\alpha$ and $\beta$. When the $\phi$ field is critical, $\xi\to\infty$, the third term becomes of order $\psi^5$, and is thus irrelevant. When $\xi$ is large but not infinite, we get the effective potential 
\begin{equation}
V_{\rm eff}(\psi)=-{\bar \alpha}\psi^2+\frac{1}{2}{\bar \beta}\psi^4+\frac{g^{3/2}\xi^{-2}}{15\pi^2\tau_0}|\psi|^3+\frac{g^{5/2}}{15\pi^2\tau_0}|\psi|^5.
\end{equation}
In addition to the $\psi^5$ term there is another term of order $\psi^3$, which may drive the second-order quantum phase transition to first order.

Let us consider a massive $\phi$ field. Carrying out the same rescaling as we made for $z=1$, we get the rescaled effective potential of the form
\begin{equation}
{\hat V}_{\rm eff}(u)=-Au^2+\frac{1}{2}Bu^4+(1+u^2)^{5/2}.
\end{equation}
For  large negative  $B$, we obtain a first-order quantum phase transition (see Fig. 5(a)).

\subsection{Fluctuations with $d=3, z=3$}
When the $\phi$ field has dynamical exponent $z=3$, e.g. for ferromagnetic fluctuations, its propagator is
\begin{equation}
\chi_0({\mathbf k},\omega_n)=\frac{1}{\gamma\frac{|\omega_n|}{k}+k^2+\xi^{-2}}.
\end{equation}
Thus the one-loop correction to the effective potential at $T=0$ is determined from
\begin{equation}
\frac{\delta V_{\rm eff}^{(1)}[\psi]}{\delta\psi}=2g\psi\int \frac{d\omega}{2\pi}\int \frac{d^d{\mathbf k}}{(2\pi)^d}\frac{1}{\gamma\frac{|\omega|}{k}+k^2+\xi^{-2}+g\psi^2},
\end{equation}
with a momentum cutoff at $|\mathbf k|=\Lambda$, and a frequency cutoff at $\gamma|\omega|=\Lambda^3$. The frequency integral gives 
\begin{equation}
\frac{\delta V_{\rm eff}^{(1)}[\psi]}{\delta\psi}=\frac{g\psi}{4\pi^4\gamma}\int_0^{\Lambda}dkk^3\ln\left[1+\frac{\Lambda^3}{k^3+k(\xi^{-2}+g\psi^2)}\right],
\end{equation}
and the momentum integral further leads to the result
\begin{equation}
V_{\rm eff}(\psi)=-{\bar \alpha}\psi^2+\frac{1}{2}{\bar \beta}\psi^4+\frac{1}{96\pi^4\gamma}(\xi^{-2}+g\psi^2)^3\ln(\xi^{-2}+g\psi^2).
\end{equation}
When $\phi$ is critical, $\xi\to\infty$, the third term is of the form $\psi^6\ln\psi$, which is irrelevant. For finite $\xi$, there is also a term of the form $\psi^4\ln\psi$, which will drive the second-order quantum phase transition to first order. 

For general $\xi$, the rescaled effective potential reads
\begin{equation}
{\hat V}_{\rm eff}(u)=-Au^2+\frac{1}{2} Bu^4+(1+u^2)^3\ln(1+u^2),
\end{equation}
which can lead to a first-order quantum phase transition (see fig. 5(b)).

\subsection{Fluctuations with $d=3, z=4$}
For a dirty metallic ferromagnet, the dynamical exponent is $z=4$. In this case, with the propagator 
\begin{equation}
\chi_0({\mathbf k},\omega_n)=\frac{1}{\gamma'\frac{|\omega_n|}{k^2}+k^2+\xi^{-2}},
\end{equation}
the rescaled effective potential reads
\begin{equation}
{\hat V}_{\rm eff}(u)=-Au^2+\frac{1}{2} Bu^4-(1+u^2)^{7/2}.
\end{equation}
Higher order terms need to be included at large $u$ to maintain stability.
When the $\phi$ field is critical, the third term is of order $\phi^7$, which is irrelevant. When the $\phi$ field is massive but light, there will also be a term of order $\phi^5$ which is again irrelevant. For general $\phi$, in order for $u=0$ to be a local minimum, we need to have $A<-7/2$. In this case, ${\hat V}'_{\rm eff}(u)=0$ has only one positive solution. Thus we have a second-order quantum phase transition.

\subsection{Fluctuations in $d=2$ and $d=1$}
We can calculate the fluctuation-induced effective potential in other dimensions in the same way as above. For $d=2, z=1$, and also for $d=1, z=2$, with the rescaled field defined by $u^2\equiv \frac{g\psi^2}{\xi^{-2}}$, the rescaled effective potential is of the form
\begin{equation}
{\hat V}_{\rm eff}(u)=-Au^2+\frac{1}{2}Bu^4-(1+u^2)^{3/2}.
\end{equation}
When the $\phi$ field is critical, the third term becomes of order $-|\psi|^3$, of the Halperin-Lubensky-Ma type, thus the quantum phase transition is first-order. Generally when $A<-1.5, AB>-0.5, B(A+B)>-0.25$, $u=0$ will be a local minimum of the rescaled effective potential ${\hat V}_{\rm eff}$, and there are two other local minima at nonzero $u$. Hence there is again a first-order quantum phase transition (see Fig. 5(c)).
Otherwise there will be a second-order phase transition.

The effective potential in the case with $d=2, z=2,$ and $d=1, z=3$ turns out be of the same form as that of $d=3, z=1$, as expected from the fact that both cases have the same effective dimension $d+z=4$. The case $d=2, z=3$ is the same as $d=3, z=2$.

For $d=1, z=1$, the effective potential takes the form 
\begin{equation}
{\hat V}_{\rm eff}(u)=-Au^2+\frac{1}{2}Bu^4-(1+u^2)\ln(1+u^2),
\end{equation}
which leads to a first-order phase transition for $B<1$ (see Fig. 5(d)).
The third term reduces to $\psi^2\ln\psi$ when $\phi$ is critical. In this case the quantum phase transition is always first order for any positive value of $B$.

\subsection{Summary of the classical + quantum cases}

In the table below, we list the most dangerous terms generated from integrating out the fluctuating fields. The second row in the table corresponds to the case where $\phi$ is critical or massless, and the third row has $\phi$ massive. 
\begin{center}
    \begin{tabular}{  | c | c | c  | c | c | c  | c |  }
    \hline
    $d+z$ & 2&3&4&5&6&7 \\ \hline
    massless & $\psi^2\ln\psi$&$\psi^3$&$\psi^4\ln\psi$&$\psi^5$&$\psi^6\ln\psi$&$\psi^7$  \\ \hline
    massive & $(\psi^2+1)\ln\psi$&$\psi^3+\psi$&$\psi^2\ln\psi$&$\psi^3$&$\psi^4\ln\psi$&$\psi^5$ \\
    \hline
    \end{tabular}
\end{center}

One can clearly see that in the massless case, the fluctuations are irrelevant when $d+z\geqslant 5$, while in the massive case, they are irrelevant for $d+z\geqslant 7$. Otherwise the second-order quantum phase transition can be driven to first order. The order of the correction is readily understood from the general structure of the integrals. With effective dimension $d+z$, in the massless case one has $\delta V/\delta\psi\sim \psi\int d^{d+z}k(1/k^2)$. Since $k^2\sim \psi^2$, this gives the correct power $\delta V\sim \psi^{d+z}$. Replacing $g\psi^2$ by $g\psi^2+\xi^{-2}$ and then carrying out the expansion in $\xi^{-2}/g\psi^2$, one gets for the massive case a reduction by $2$ in the power. We also notice the even/odd effect in the effective potential: for $d+z$ even, there are logarithmic corrections. The case $d+z=4$ can be easily understood, as the system is in the upper critical dimension, and logarithmic corrections are expected. We still do not have a simple intuitive understanding of the logarithm for $d+z=2, 6$.

\section{Two fluctuating fields}
We consider in this section the case where the two coupled quantum fields are both fluctuating substantially. 
The partition function now becomes
\begin{equation}
Z=\int {\cal D}\psi({\mathbf r},\tau)\int {\cal D}\phi({\mathbf r},\tau)  \exp\left({-S_{\psi}-S_{\phi}-S_{\psi\phi}}\right).
\end{equation}
We will use RG equations to determine the phase diagram of this system.  When there is no stable fixed point, or the initial parameters lie outside the basin of attraction of the stable fixed points, the flow trajectories will show runaway behavior, which implies a first-order phase transition \cite{Fisher77, Lubensky78, Rudnick78, Amit81, Cardy96}. The spin-density-wave transitions in some cuprates and pnictides fall in this category \cite{Tranquada95, Ando02, Tailefer07, Kivelson98, Fradkin09, Dai08, Xu08, Fang08, Huang08, Krellner08, Yan08, WangDai08, McQueeney08, Vicari06}.

\begin{figure}[ht]
\begin{center}
\includegraphics[width=6cm, clip]{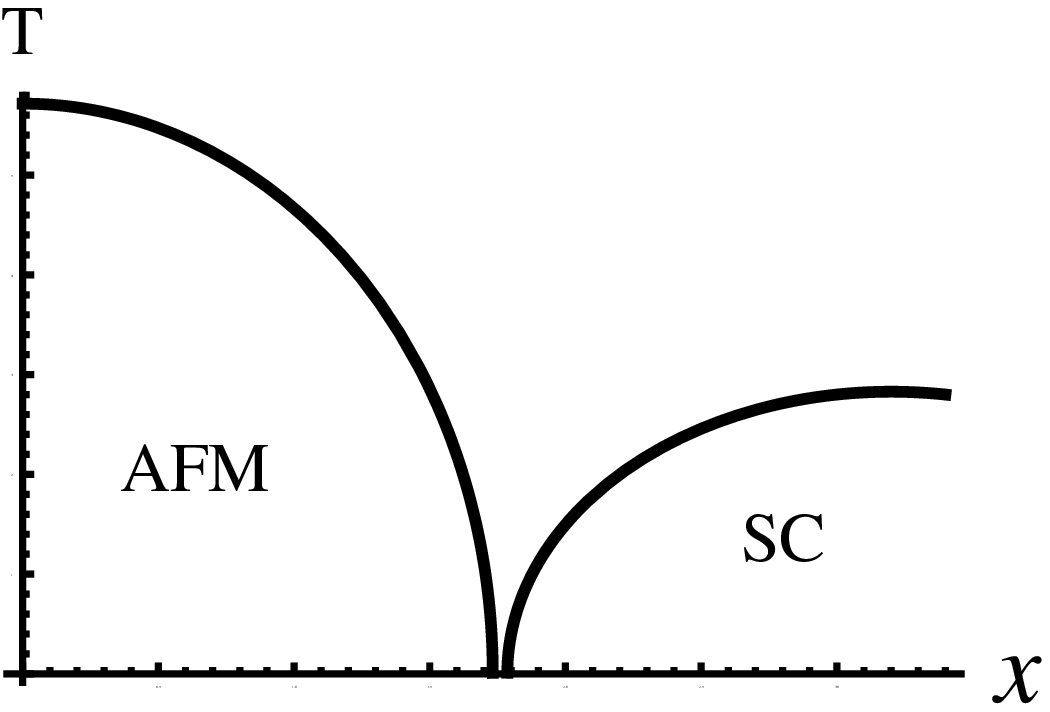}
\includegraphics[width=6cm, clip]{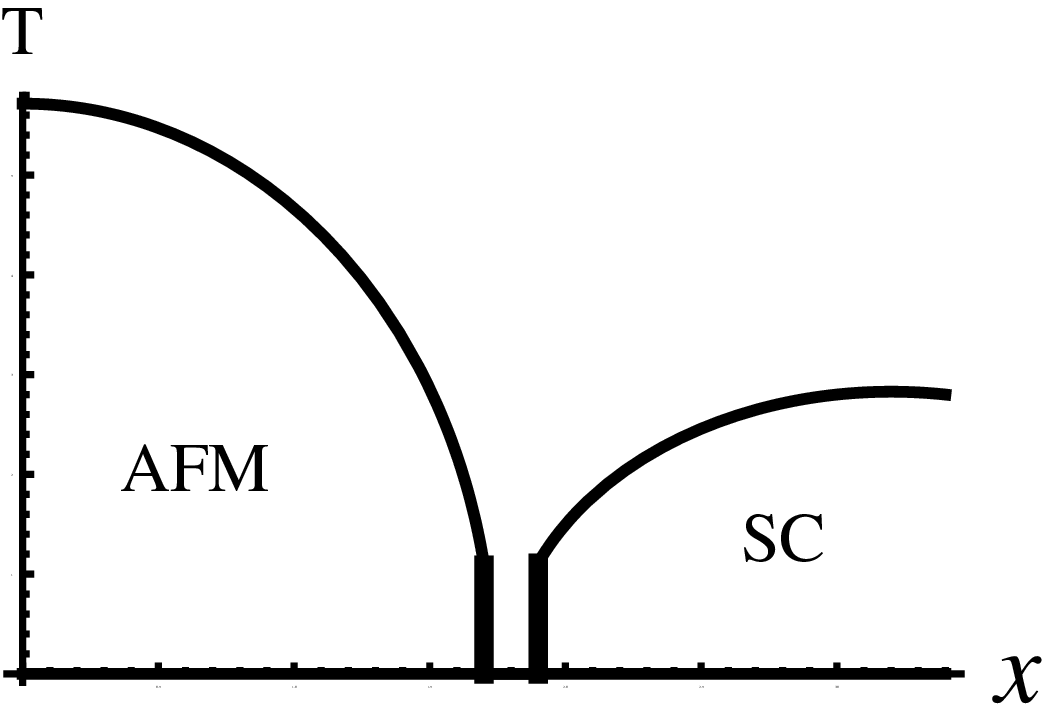}
\end{center}
\caption{Illustration of the fluctuation-induced first-order phase transition in the case of two quantum fields. Here for concreteness we consider the antiferromagnetic order and superconducting order. At low temperatures, the phase transitions may become first order (the thick vertical lines), due to fluctuations.}
\label{fig3}
\end{figure}

We have considered in the previous sections coupling two single component fields, having in mind that this simplified model captures the main physics of competing orders. However, we will see below that when the quantum fluctuations of both fields are taken into account, the number of components of the order parameters do play important roles. So from now on we consider explicitly a $n_1$-component vector field $\boldsymbol \psi$ and a $n_2$-component vector field $\boldsymbol \phi$ coupled together. When both fields have dynamical exponent $z=1$, the action reads
\begin{equation}
 \begin{split}
S_{\psi}=&\int d^d{\mathbf r}d\tau\left[-\alpha_1 |{\boldsymbol \psi}|^2+\frac{1}{2}\beta_1 |{\boldsymbol \psi}|^4+\frac{1}{2}|\partial_\mu \boldsymbol \psi|^2\right],
\\S_{\phi}=&\int d^d{\mathbf r}d\tau\left[-\alpha_2 |\boldsymbol \phi|^2+\frac{1}{2}\beta_2 |\boldsymbol \phi|^4+\frac{1}{2}|\partial_\mu \boldsymbol \phi|^2\right],
\\S_{\psi\phi}=&g \int d^d{\mathbf r}d\tau  |\boldsymbol\psi|^2 |\boldsymbol \phi|^2,
\end{split}
\label{action1}
\end{equation}
where $\mu=0, 1, \cdots, d$. 
This quantum mechanical problem is equivalent to a classical problem in one higher dimension. 
Then one can follow the standard procedure of RG: first decompose the action into the fast-moving part, the slow-moving part and the coupling between them. The Green's functions are $G_{\psi}=1/(-2\alpha_1+k^2+\omega^2)$ and $G_{\phi}=1/(-2\alpha_2+k^2+\omega^2)$. The relevant vertices are $\beta_1\psi_s^2\psi_f^2, \beta_2\phi_s^2\phi_f^2, g\psi_s^2\phi_f^2, g\psi_f^2\phi_s^2, g\psi_s\psi_f\phi_s\phi_f$. To simplify the notation we rescale the momentum and frequency according to ${\mathbf k}\to {\mathbf k}/\Lambda, \omega\to\omega/\Lambda$, so that they lie in the interval $[0, 1]$. The control parameters and couplings are rescaled according to $\alpha_{1,2}\to\alpha_{1,2}\Lambda^2$, $\beta_{1,2}\to\beta_{1,2}\Lambda^{3-d}, g\to g\Lambda^{3-d}$. Afterwards we integrate out the fast modes with the rescaled momentum and frequency in the range $[b^{-1},1]$. Finally, we rescale the momentum and frequency back to the interval $[0, 1]$, thus ${\mathbf k}\to b{\mathbf k}, \omega\to b\omega$, and the fields are rescaled accordingly with $\psi\to b^{(d-1)/2}\psi, \phi\to b^{(d-1)/2}\phi$.
Using an $\epsilon$-expansion, where $\epsilon=3-d$, one obtains the set of RG equations to one-loop order,
\begin{equation}
 \begin{split}
\frac{d\alpha_i}{dl}=&2\alpha_i-\frac{1}{8\pi^2}[(n_i+2)\beta_i(1+2\alpha_i)+n_jg(1+2\alpha_j)],
\\ \frac{d\beta_i}{dl}=&\epsilon\beta_i-\frac{1}{4\pi^2}[(n_i+8)\beta_i^2+n_jg^2],
\\ \frac{dg}{dl}=&g\left(\epsilon-\frac{1}{4\pi^2}\left[(n_1+2)\beta_1+(n_2+2)\beta_2+4g\right]\right),
\end{split}
\label{rg1}
\end{equation}
with index $i, j=1, 2$, and $i\neq j$. These equations are actually more general than considered above. They also apply to generic models where two fields with the same dynamical exponent $z$ are coupled together. Generally  one has $\epsilon=4-d-z$, thus a quantum mechanical model with dynamical exponent $z$ is equivalent to a classical model in dimension $d+z$.

\begin{figure}[ht]
\begin{center}
\includegraphics[width=7cm, clip]{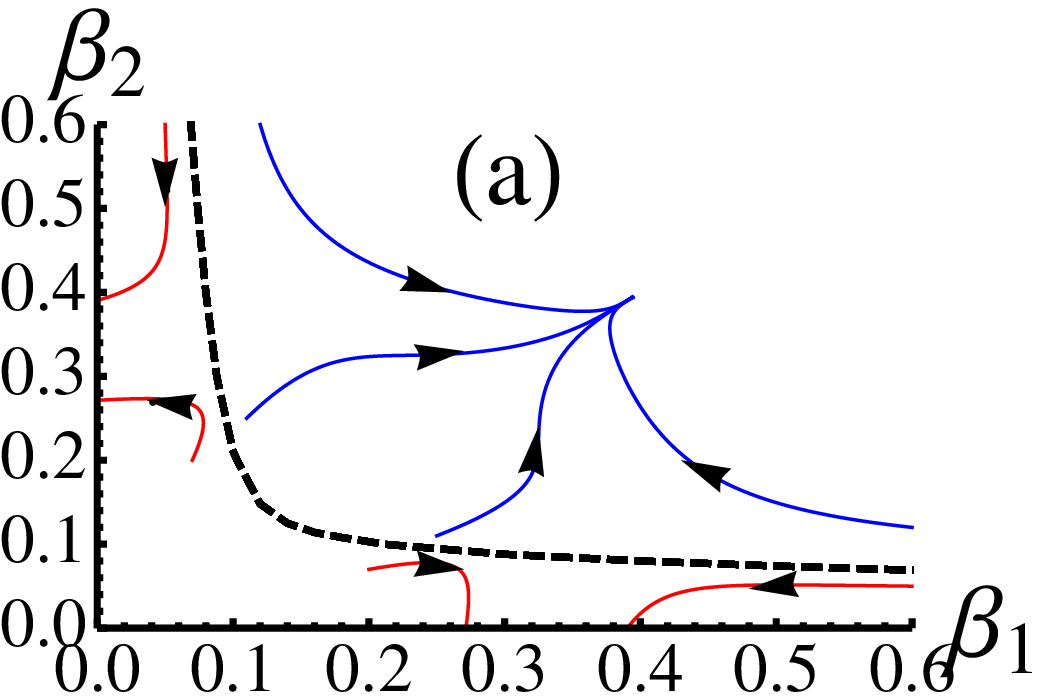}
\includegraphics[width=7cm, clip]{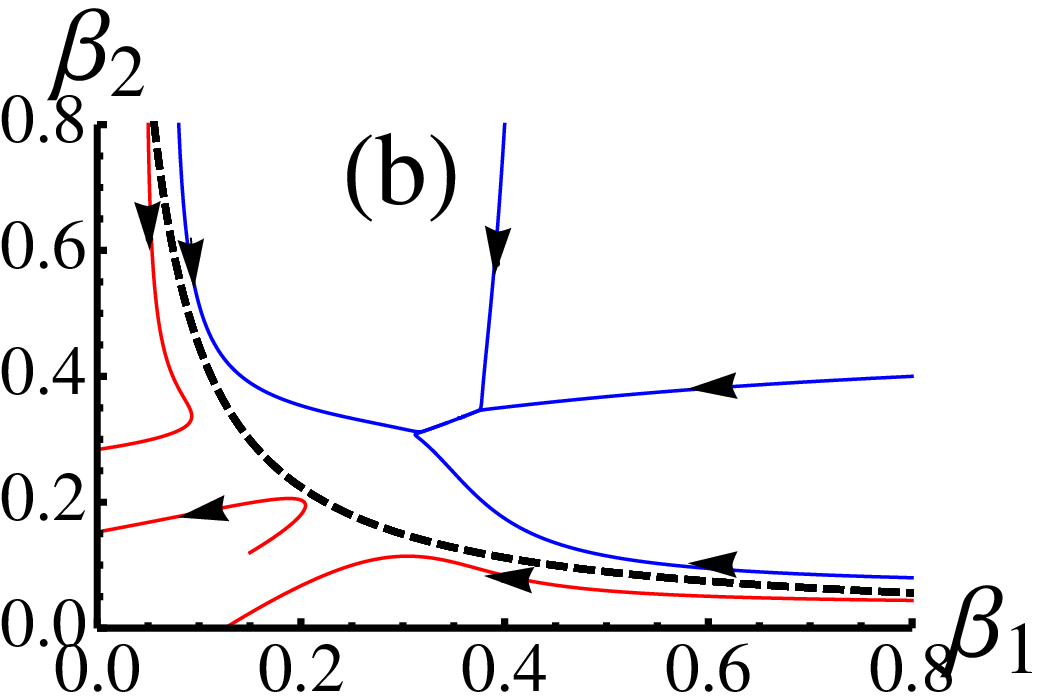}
\end{center}
\caption{(Color online) Plot of the RG trajectories in the $\beta_1-\beta_2$ plane for two quantum fields with the same dynamical exponent below the upper critical dimension. Here we have chosen $\epsilon=4-d-z=0.1$. The RG trajectories have been projected onto a constant $g$ plane with $g=g^*$, and $g^*$ the value of the coupling strength at the stable fixed point. (a) corresponds to the case $n_1=n_2=1$, where the fixed point is at $\beta_1^*=\beta_2^*=g^*= 4\pi^2\epsilon/(n_1+n_2+8)\simeq 0.3948$. (b) corresponds to the case $n_1=2, n_2=3$, where the fixed point is at $(\beta_1^*, \beta_2^*, g^*)=4\pi^2\epsilon(0.0905, 0.0847, 0.0536)\simeq (0.3573, 0.3344, 0.2116)$. In both cases we found that, above some curve (the dashed lines), the RG trajectories flow to the corresponding stable fixed point, while below this curve,  the RG trajectories show runaway behavior.}
\label{fig9}
\end{figure}

It is known that the above equations have six fixed points \cite{Fisher76}, four of which have the two fields decoupled, i.e.,  $g^*=0$. They are the Gaussian-Gaussian point at $(\beta_1^*, \beta_2^*)=(0, 0)$, the Heisenberg-Gaussian point at $(\beta_1^*, \beta_2^*)=(4\pi^2\epsilon/(n_1+8), 0)$, the Gaussian-Heisenberg point at $(\beta_1^*, \beta_2^*)=(0, 4\pi^2\epsilon/(n_2+8))$, and the decoupled Heisenberg-Heisenberg point at $(\beta_1^*, \beta_2^*)=(4\pi^2\epsilon/(n_1+8), 4\pi^2\epsilon/(n_2+8))$. The isotropic Heisenberg fixed point is at $\beta_1^*=\beta_2^*=g^*= 4\pi^2\epsilon/(n_1+n_2+8), \alpha_1^*=\alpha_2^*=\epsilon(n_1+n_2+2)/4(n_1+n_2+8)$. Finally there is the biconical fixed point with generally unequal values of $\beta_1^*$, $\beta_2^*$, and $g^*$. In the case, with $n_1=n_2=1$, this is at $(\beta_1^*, \beta_2^*, g^*)=2\pi^2\epsilon/9(1,1,3)$. For $n_1=2, n_2=3$, one has $(\beta_1^*, \beta_2^*, g^*)=4\pi^2\epsilon(0.0905, 0.0847, 0.0536)$.

We find that there is always just one stable fixed point for $d+z<4$, below the upper critical dimension \cite{Fisher76}. The isotropic Heisenberg fixed point is stable when $n_1+n_2<n_c=4-2\epsilon+O(\epsilon^2)$, the biconical fixed point is stable when $n_c<n_1+n_2<16-n_1n_2/2+O(\epsilon)$, and when $n_1n_2+2(n_1+n_2)>32+O(\epsilon)$, the decoupled Heisenberg-Heisenberg point is the stable one. When the initial parameters are not in the basin of attraction of the stable fixed point, one obtains runaway flow, strongly suggestive of a first-order phase transition. Consider for example $n_1=2, n_2=3$, where the biconical fixed point is stable. For two critical points not too separated, that is, $|\alpha_1-\alpha_2|$ not too large, when $g>\sqrt{\beta_1\beta_2}$ the RG flow shows runaway behavior, and one gets a first-order quantum phase transition. The corresponding classical problem has  been discussed in \cite{Nagaosa00}. We notice the difference from the case with two competing classical fields, where one also obtains the same condition for the couplings $\gamma>\sqrt{\beta_1\beta_2}$ in order to have a first-order phase transition. There, the two ordered phases are required to overlap in the absence of the coupling, in other words, one needs to have $x_1<x_2$. However, in the quantum mechanical case we are considering here, this is not necessary. We plot in Fig. 7 the RG trajectories for two cases (a) $n_1=n_2=1$ and (b) $n_1=2, n_3=3$, where in both cases, below some curve, runaway behavior in the RG trajectories is found.

\begin{figure}[ht]
\begin{center}
\includegraphics[width=7cm, clip]{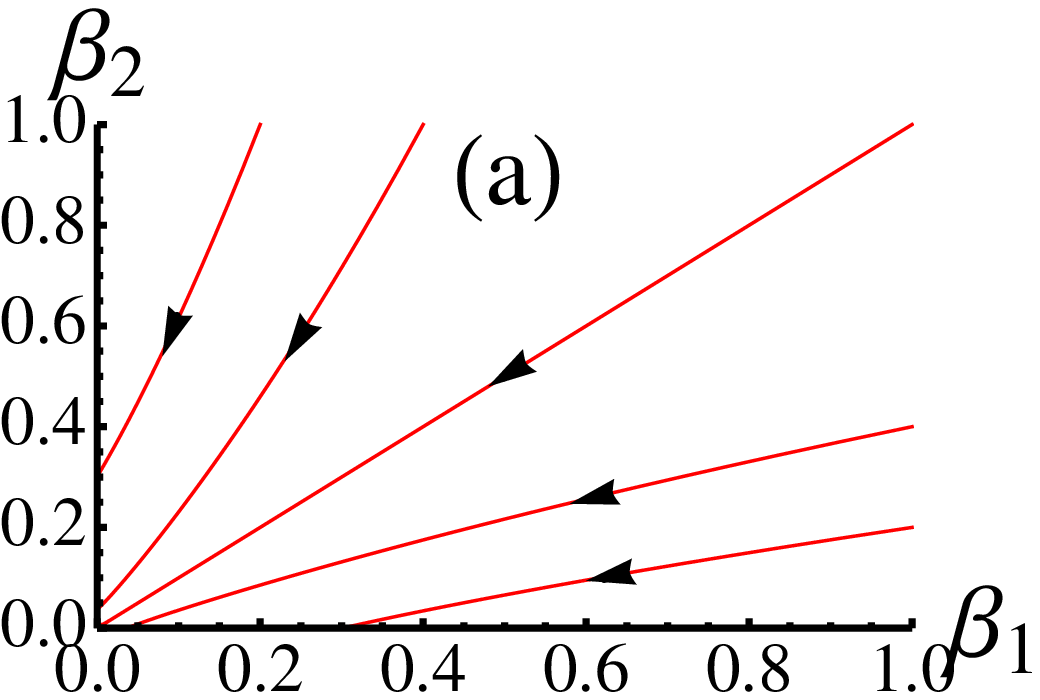}
\includegraphics[width=7cm, clip]{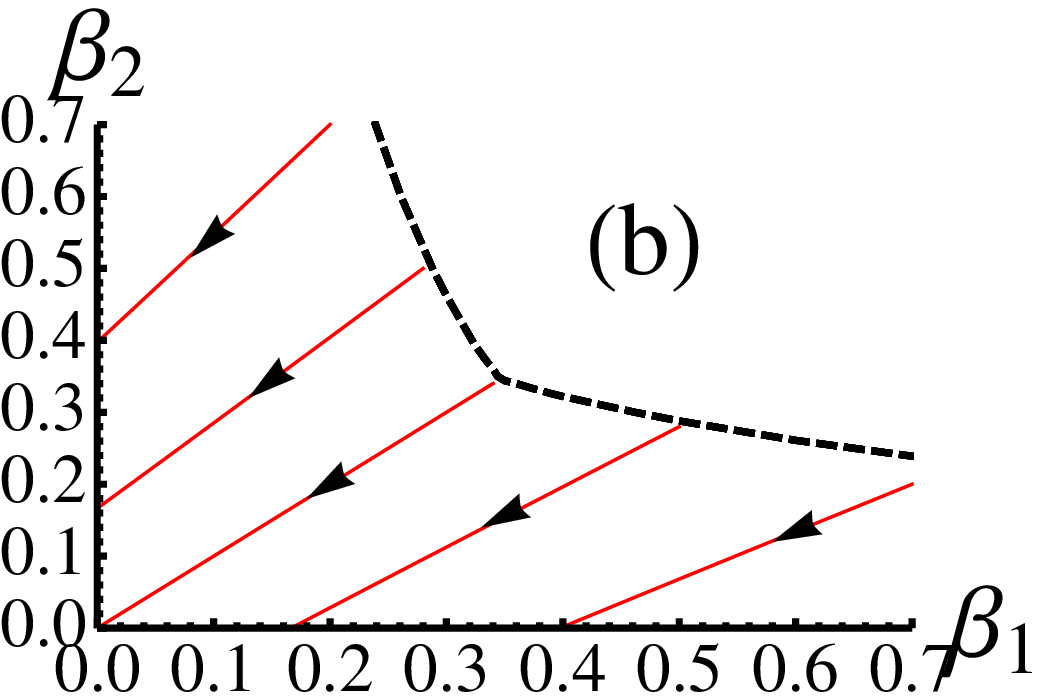}
\end{center}
\caption{(Color online) Plot of the RG trajectories in the $\beta_1-\beta_2$ plane for two quantum fields with the same dynamical exponent in and above the upper critical dimension. The RG trajectories have been projected to a constant $g$ plane. And we have chosen $n_1=n_2=3$. (a) corresponds to the case exactly at the critical dimension with $\epsilon=4-d-z=0$. In this case there is only one fixed point with $\beta_1^*=\beta_2^*=g^*=0$, the Gaussian fixed point, which is unstable. We found runaway flows everywhere. (b) corresponds to the case above the critical dimension, where the Gaussian fixed point is the stable one. Here we have chosen $\epsilon=4-d-z=-0.1$. We found, below some curve (the dashed line), that the RG trajectories show runaway behavior.}
\label{fig10}
\end{figure}

When $d+z=4$, all the other fixed points coalesce with the Gaussian point, forming an unstable fixed point, thus leading to a first-order phase transition (see Fig. 8(a) ). A similar model with an extra coupling and $n_1=n_2=3$ has been discussed by Qi And Xu \cite{Qi09}, where runaway flows were also identified. Another similar problem with $d=2, z=2$ was studied by Millis recently \cite{Millis10}, where a fluctuation-induced first-order quantum phase transition was shown to occur. We also notice that in some situations, including fluctuations of the order parameter itself may drive the supposed-to-be first-order transitions to second order for both classical and quantum phase transitions \cite{Parisi81, Belitz02, Jakubczyk09, Metzner0902}. 

For $d+z>4$, the stabilities are interchanged. The Gaussian fixed point becomes the most stable one. So the basin of attraction of the stable fixed point changes. We found numerically that for a given coupling strength $g$, in the $\beta_1-\beta_2$ plane, the RG trajectories show runaway behavior when the initial points lie below some curve (see Fig. 8(b)). That is, when the coupling between the two fields is strong enough, the QPTs become first order. Just above these curves, we found that the RG trajectories will enter the domain with negative $\beta_1$ or negative $\beta_2$, and then converge to the Gaussian fixed point. For $\beta_1, \beta_2$ large enough, the RG trajectories just converge to the Gaussian fixed point without entering the negative domain.

\subsection{Competing orders with different dynamical exponents}
We consider next coupling a $z=1$ field to another field with dynamical exponent $z=z_1\geqslant 2$. To our knowledge, such models of two competing order parameters with different dynamical exponents have not been studied previously. The action now takes the form
\begin{equation}
 \begin{split}
S_{\psi}=&\int d^d{\mathbf k}d\omega \left( -\alpha_1+\frac{k^2}{2}+\frac{\gamma_1}{2}\frac{|\omega|}{k^{z_1-2}} \right)|{\boldsymbol \psi}|^2+\int d^d{\mathbf r}d\tau\frac{1}{2}\beta_1 |{\boldsymbol \psi}|^4,
\\S_{\phi}=&\int d^d{\mathbf r}d\tau\left[-\alpha_2 |\boldsymbol \phi|^2+\frac{1}{2}\beta_2 |\boldsymbol \phi|^4+\frac{1}{2}|\partial_\mu \boldsymbol \phi|^2\right],
\\S_{\psi\phi}=&g \int d^d{\mathbf r}d\tau  |\boldsymbol\psi|^2 |\boldsymbol \phi|^2.
\end{split}
\label{action2}
\end{equation}
The new parameter $\gamma_1$ has dimension $[\gamma_1]=L^{1-z}$, and its one-loop RG equation is simply
\begin{equation}
\frac{d\gamma_1}{dl}=(z-1)\gamma_1.
\label{rgg}
\end{equation}
The Green's function for the $\psi$ field becomes $G_{\psi}=1/(-2\alpha_1+k^2+\gamma_1|\omega|/k^{z_1-2})$. The RG equations for the other parameters are modified accordingly,
\begin{equation}
 \begin{split}
\frac{d\alpha_1}{dl}=&2\alpha_1-\frac{\Omega_{d}}{\pi\gamma_1}(n_1+2)\beta_1(\ln 2+2\alpha_1)-\Omega_{d+1}n_2g(2+2\alpha_2),
\\ \frac{d\alpha_2}{dl}=&2\alpha_2-\Omega_{d+1}(n_2+2)\beta_2(2+2\alpha_2)-\frac{\Omega_{d}}{\pi\gamma_1}n_1g(\ln 2+2\alpha_1),
\\ \frac{d\beta_1}{dl}=&\epsilon\beta_1-\frac{2\Omega_{d}}{\pi\gamma_1}(n_1+8)\beta_1^2-2\Omega_{d+1}n_2g^2,
\\ \frac{d\beta_2}{dl}=&\epsilon\beta_2-2\Omega_{d+1}(n_2+8)\beta_2^2-\frac{2\Omega_{d}}{\pi\gamma_1}n_1g^2,
\\ \frac{dg}{dl}=&g\left(\epsilon-\frac{2\Omega_{d}}{\pi\gamma_1}(n_1+2)\beta_1-2\Omega_{d+1}(n_2+2)\beta_2-8\frac{\Omega_d}{2\pi}\frac{ 2 \gamma_1\ln{\gamma_1} +\pi  }{1+\gamma_1^2}
g\right),
\end{split}
\label{rgab}
\end{equation}
where $\epsilon=3-d$ and $\Omega_{d}=2\pi^{d/2}/(2\pi)^{d}\Gamma[d/2]$ is the volume of the $d$-dimensional unit sphere. The derivation of the above RG equations is included in the appendix. We notice from the above procedure that when the two fields have the same dynamical exponent $z>1$, one can rescale the couplings to ${\tilde\beta}_1=\beta_1/\gamma, {\tilde\beta}_2=\beta_2/\gamma, {\tilde g}=g/\gamma$, and these new parameters satisfy the RG equations (\ref{rg1}) with ${\tilde\epsilon}=4-d-z$.

The presence of two different dynamical exponents obviously complicates the problem. It is generally expected that the modes with a larger dynamical exponent dominates the specific heat of the system, since they have a large phase space, while the modes with a smaller dynamical exponent may produce infrared singularities, since they have a smaller upper critical dimension \cite{Peter09}. In the absence of the coupling between the two fields, we have the RG equations 
\begin{equation}
 \begin{split}
\\ \frac{d\tilde{\beta}_1}{dl}=&(4-d-z)\tilde{\beta}_1-\frac{2\Omega_{d}}{\pi}(n_1+8)\tilde{\beta}_1^2,
\\ \frac{d\beta_2}{dl}=&(3-d)\beta_2-2\Omega_{d+1}(n_2+8)\beta_2^2.
\end{split}
\label{rg3}
\end{equation}
For $d=3$, $\beta_2$ is marginal with an unstable fixed point, while $\tilde{\beta}_1$ is irrelevant and its Gaussian fixed point is stable.

\begin{figure}[ht]
\begin{center}
\includegraphics[width=7cm, clip]{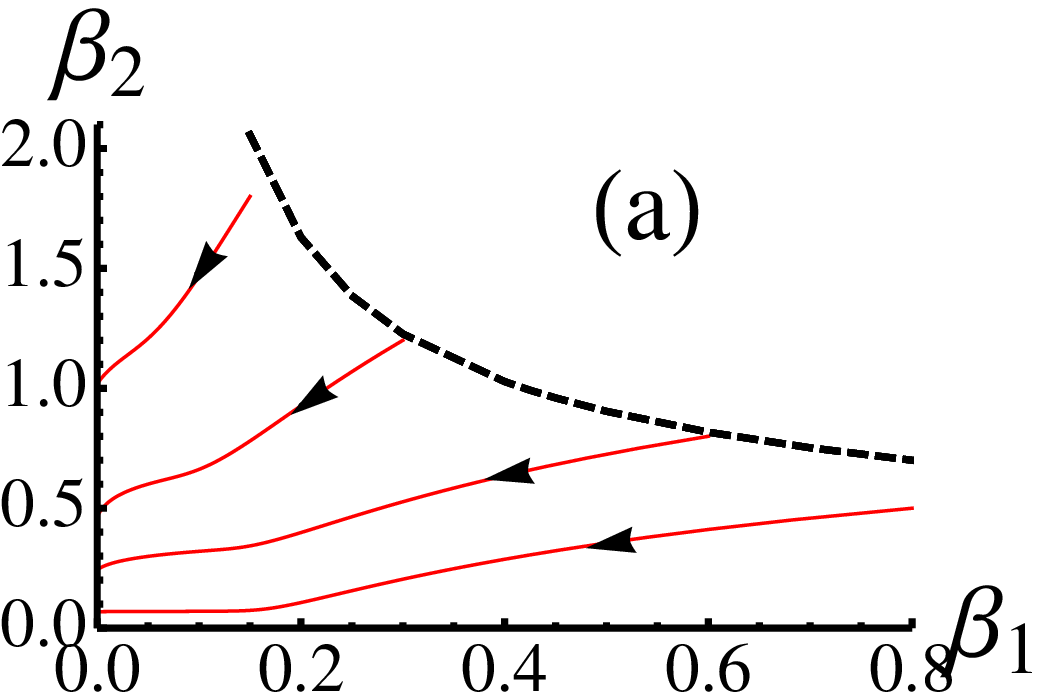}
\includegraphics[width=7cm, clip]{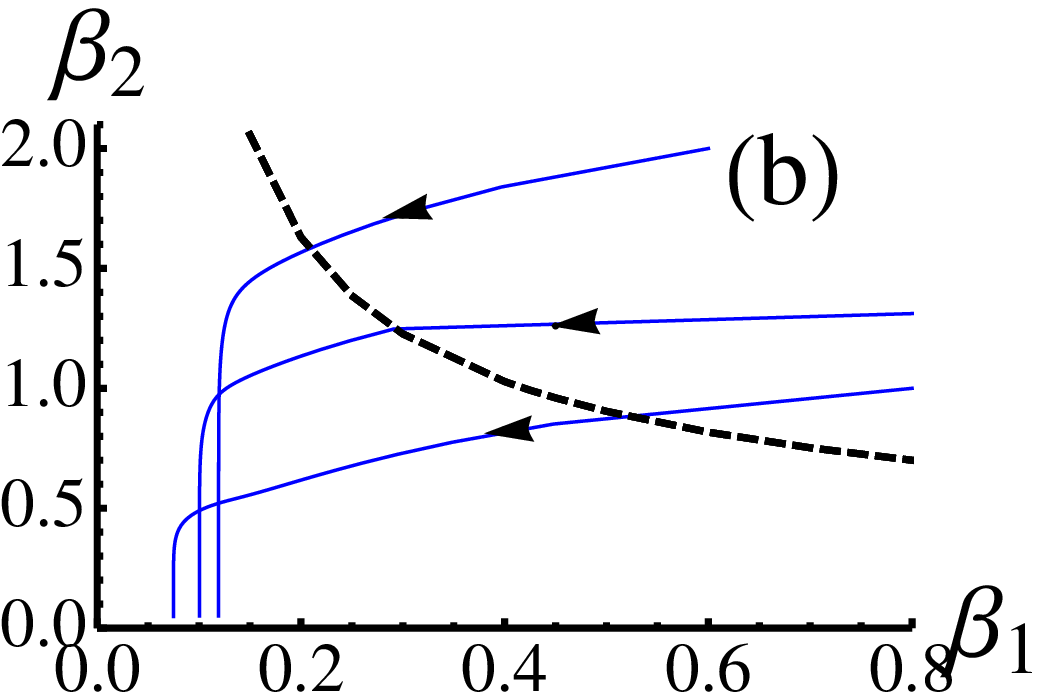}
\end{center}
\caption{(Color online) Plot of the RG trajectories in the $\beta_1-\beta_2$ plane for two coupled quantum fields with different dynamical exponents. The RG trajectories have been projected to a constant $g$ plane with $g=1$. We have chosen the spatial dimension to be $d=3$, the dynamical exponents $z_1=2,  z_2=1$ and the number of field components $n_1=n_2=3$. (a) shows the RG trajectories originating from the region below the dashed line, which flow to negative $\beta_1$ or negative $\beta_2$ regions. (b) shows the RG trajectories originating from the region above the dashed line, and those flow to the stable points on the positive axes of $\beta_1$, the location of which is sensitive to the initial value of the parameters..}
\label{fig11}
\end{figure}

Generally, for $z>1$, if the initial value of $\gamma_1$ is nonzero, the absolute value of $\gamma_1$ will increase exponentially. The RG equation for $\beta_2$ becomes independent of other parameters,
\begin{equation} 
 \frac{d\beta_2}{dl}=\epsilon\beta_2-2\Omega_{d+1}(n_2+8)\beta_2^2.
\end{equation}
We are interested in the case $\epsilon=0$, for which $\beta_2$ is readily solved to be
\begin{equation} 
 \beta_2(l)=\frac{1}{\bar{\beta}_2^{-1}+ 2\Omega_{4}(n_2+8)(l-l_{\rm cr})},
\end{equation}
with $\bar{\beta}_2$ taken at the crossover scale $l_{\rm cr}$ at which the $\beta_2^2$ term begins to dominate the $g^2$ term. Only the sign of $\bar{\beta}_2$ matters. If $\bar{\beta}_2>0$, as $l$ increases, $\beta_2$ will decay to zero, flowing to its Gaussian fixed point. From the simplified RG equations for $g$,
\begin{equation}
 \\ \frac{dg}{dl}=-2\Omega_4(n_2+2)g\beta_2,
\label{rg22}
\end{equation}
one can see that with a lower power in $\beta_2$, $g$ drops to zero even more quickly than $\beta_2$. Taking $\beta_2$ as quasi-static when considering the evolution of $g$, one notices that $g$ decays exponentially as $g(l)\sim \exp(-2\Omega_4(n_2+2)\beta_2l)$. So $d\beta_1/dl$ also decays exponentially, and before $\beta_2$ goes to zero, $\beta_1$ already stabilizes to a finite value $\beta_1^*$, which depends on the initial value of $\beta_1$. Actually from the simplified RG equations for $\beta_1, \beta_2, g$ with $1/\gamma_1$ set to zero, one can see directly that the fixed points are at $\beta_2^*=g^*=0$, with $\beta_1^*$ any real number: we have a line of fixed points. When $\beta_1^*>0$, there will be a second-order phase transition. When $\beta_1^*<0$, the transition becomes first order (see Fig. 9).

 If $\bar{\beta}_2<0$, the absolute value of $\beta_2$ will increase without bound. Subsequently $g$ and $\beta_1$ also diverge, leading to runaway flows.

\section{Conclusion}

Quantum criticality in the presence of competing interactions is an important guiding concept that allows us to organize a framework for emergent states near QCPs. Here we investigated the 
stability of a quantum critical point in the presence of competing orders. We focused on a simple quadratic-quadratic interaction, where coupling between two competing phases is assumed to be of $g \psi^2 \phi^2$ form. We find that QCPs are often unstable and transform into first order lines of transitions. The detailed scenario on how the instability develops depends on the precise nature of the competing interactions, dynamical exponents and strength of the coupling. The general trend we observe is that competing interactions, be they classical or quantum, often lead to the instability of QCPs. This instability in fact always occurs, in the cases we have investigated, if the coupling $g$ is strong enough. We thus conclude that  breakdown of QCPs is a ubiquitous phenomenon. The magnitude of the specific heat jump in some first order transitions (the \emph{classical} + \emph{classical} case) is of the same order as the specific heat released in a second order transition and these first order transitions are strong, and not weakly first order as found in Halperin-Lubensky-Ma. An immediate consequence of this breakdown is that we can expect spatially modulated inhomogeneous phases to be present near QCPs, given their propensity to turn into first order transitions. The wide likelihood identified here of first order transitions preempting a QCP leads us to anticipate the nucleation and metastability phenomena associated with such transitions \cite{Hohenberg95}. Additionally, proximity to first-order transitions makes auxiliary fields (e. g. magnetic field, strain) and disorder very important over substantial parameter regions \cite{Alan93}.

The broad similarities we pointed out between QCPs and AdS/CFT models offers an interesting possibility that in fact AdS models are also spatially inhomogeneous. More detailed analysis that allows breakdown of scaling,  specific for AdS/CFT is suggested. 

We derived the renormalization group equations for two coupled order parameters with different dynamical exponents. We found that there are a line of fixed points, which is quite different from the case where two order parameters have the same dynamical exponent. Very recently, there have appeared some interesting reports \cite{Peter09, Chubukov10} investigating the effects of the presence of two order parameters with different dynamical exponents near the Pomeranchuk instability \cite{Pomeranchuk58}, as examples of multiscale quantum criticality. It would be interesting to see how the presence of two different dynamical exponents, and the coupling between the corresponding order parameters, affect the scaling of resistivity, especially whether a linear-resistivity is possible, overcoming the "no-go" theorem for single parameter scaling \cite{Phillips05}. 

In this paper, we have confined ourselves to the framework of Hertz-Millis-Moriya \cite{Hertz76, Millis93, Moriya85}, considering only the interplay of bosonic order parameters. It would also be interesting to study the electronic instabilities, to see whether the superconducting instabilities and Pomeranchuk instabilities are enhanced in fermionic quantum critical states. Fermi liquids, even with repulsive interactions, are unstable towards forming a superconducting state, due to the Kohn-Luttinger effect \cite{Kohn65} resulting from the presence of a sharp Fermi surface. For the fermionic quantum critical states, the momentum distribution function may have only higher order singularities \cite{Senthil08}. It would be interesting to check whether the Kohn-Luttinger effect is still active in this case.

\section{Acknowledgments}

We are grateful to  G. Aeppli, E. D. Bauer, Y. Dubi, P. Littlewood, F. Ronning, T. Rosenbaum and P. W$\rm\ddot{o}$lfle for stimulating discussions over the years about stability of QCPs. J. S. and J. Z. thank K. Schalm and V. Juricic for helpful discussions on AdS/CFT correspondence and the RG results. This work was supported by US DoE, BES and LDRD. J. S. and J. Z. are supported by the Nederlandse Organisatie voor Wetenschappelijk Onderzoek (NWO) via a Spinoza grant.

\section{Appendix: Derivation of the RG equations for two fields with different dynamical exponents}

In this appendix, we will derive the RG equations of two competing orders with different dynamical exponents. We follow the notation of \cite{Altland}.
Our starting point is the action (\ref{action2}). First we count the dimensions of the field operators and all the parameters:
 \begin{equation}
 \begin{split}
\\ [r]=[\tau]=&L,
\\ [k]=[\omega]=&L^{-1},
\\ [\psi]=[\phi]=&L^{(1-d)/2},
\\ [\alpha_1]=[\alpha_2]=&L^{-2},
\\ [\beta_1]=[\beta_2]=&L^{d-3},
\\ [g]=&L^{d-3},
\\ [\gamma_1]=&L^{1-z}.
\end{split}
\label{dim1}
\end{equation}
Then we decompose the action into slow and fast modes. The action for the slow modes reads
\begin{equation}
 \begin{split}
\\ S^{(s)}=&S^{(s)}_{\psi}+S^{(s)}_{\phi}+S^{(s)}_{\psi\phi},
\\S^{(s)}_{\psi}=&\int d^d{\mathbf k}d\omega \left( -\alpha_1+\frac{k^2}{2}+\frac{\gamma_1}{2}\frac{|\omega|}{k^{z_1-2}} \right)|{\boldsymbol \psi}_s|^2+\int d^d{\mathbf r}d\tau\frac{1}{2}\beta_1 |{\boldsymbol \psi}_s|^4,
\\S^{(s)}_{\phi}=&\int d^d{\mathbf r}d\tau\left[-\alpha_2 |\boldsymbol \phi_s|^2+\frac{1}{2}\beta_2 |\boldsymbol \phi_s|^4+\frac{1}{2}|\partial_\mu {\boldsymbol \phi}_s|^2\right],
\\S^{(s)}_{\psi\phi}=&g \int d^d{\mathbf r}d\tau  |\boldsymbol\psi_s|^2 |\boldsymbol \phi_s|^2.
\end{split}
\label{actions1}
\end{equation}
Since we will only consider RG to one-loop order, the interaction terms in the fast modes, the contraction of which leads to second-order diagrams, can be ignored. Thus we obtain the action for the fast modes,
\begin{equation}
 \begin{split}
\\ S^{(f)}=&S^{(f)}_{\psi}+S^{(f)}_{\phi},
\\S^{(f)}_{\psi}=&\int d^d{\mathbf k}d\omega \left( -\alpha_1+\frac{k^2}{2}+\frac{\gamma_1}{2}\frac{|\omega|}{k^{z_1-2}} \right)|{\boldsymbol \psi}_f|^2,
\\S^{(f)}_{\phi}=&\int d^d{\mathbf r}d\tau\left[-\alpha_2 |\boldsymbol \phi_f|^2+\frac{1}{2}|\partial_\mu {\boldsymbol \phi}_f|^2\right],
\end{split}
\label{actions2}
\end{equation}
from which one can easily identify the Green's functions as
 \begin{equation}
 \begin{split}
\\ G^{f}_{ij}[\psi]=&\frac{\delta_{ij}}{-2\alpha_1+k^2+\gamma_1\frac{|\omega|}{k^{z_1-2}}},
\\ G^{f}_{ij}[\phi]=&\frac{\delta_{ij}}{-2\alpha_2+k^2+\omega^2}.
\end{split}
\label{green2}
\end{equation}
The coupling between the slow modes and fast modes takes the form
\begin{equation}
S_{c}=\int d^d{\mathbf r}d\tau\left[ \sum_{ijkl}F_{ijkl}\left(3\beta_1  \psi_f^i  \psi_f^j \psi_s^k \psi_s^l+3\beta_2  \phi_f^i  \phi_f^j \phi_s^k \phi_s^l \right) +g |\boldsymbol \psi_s|^2  |\boldsymbol \phi_f|^2 +g |\boldsymbol \psi_f|^2  |\boldsymbol \phi_s|^2  +4g (\boldsymbol \psi_s\cdot\boldsymbol \psi_f )  (\boldsymbol\phi_s\cdot\boldsymbol \phi_f)     \right],
\label{actioncoup}
\end{equation}
with the tensor $F_{ijkl}=\frac{1}{3}(\delta_{ij}\delta_{kl}+\delta_{ik}\delta_{jl}+\delta_{il}\delta_{jk})$.

Now we can integrate out the fast modes and see how the different parameters change accordingly. The effective action of the slow modes is determined by
\begin{equation}
\exp\left[-{S^{(s)}_{\rm eff}}\right]=\exp\left[-S^{(s)}\right]\exp\left[ -\langle S_c\rangle_f +\frac{1}{2}\langle S_c^2\rangle_f^{\rm con}    \right].
\end{equation}
In the $S_c^2$ term we take a connected average, thus the superscript "$\rm con$". 
The coefficients in the RG equations will depend on the different renormalization schemes. Here we will use the procedure that is most convenient for the problem at hand, similar in spirit to what was outlined in \cite{Herbut07}. We integrate over the momentum interval $\Lambda/b<k<\Lambda$, which after rescaling $k\to k/\Lambda$, gives $b^{-1}<k<1$. The frequency part is more complicated. We will introduce a cutoff when it is necessary, otherwise just integrate over the whole range
$-\infty<\omega<\infty$. The main reason for us to choose this RG scheme is that in calculating the third correction to the coupling $g$, the two internal lines come from order parameters with different dynamical exponents, thus the two frequencies scale differently with momentum, and this RG scheme offers a simple and self-consistent treatment of the cutoffs.

$\gamma_1$ does not receive corrections up to first-order.

\subsection{First Order Corrections to $\alpha_1$}

\begin{figure}[ht]
\begin{center}
\includegraphics[width=4cm, clip]{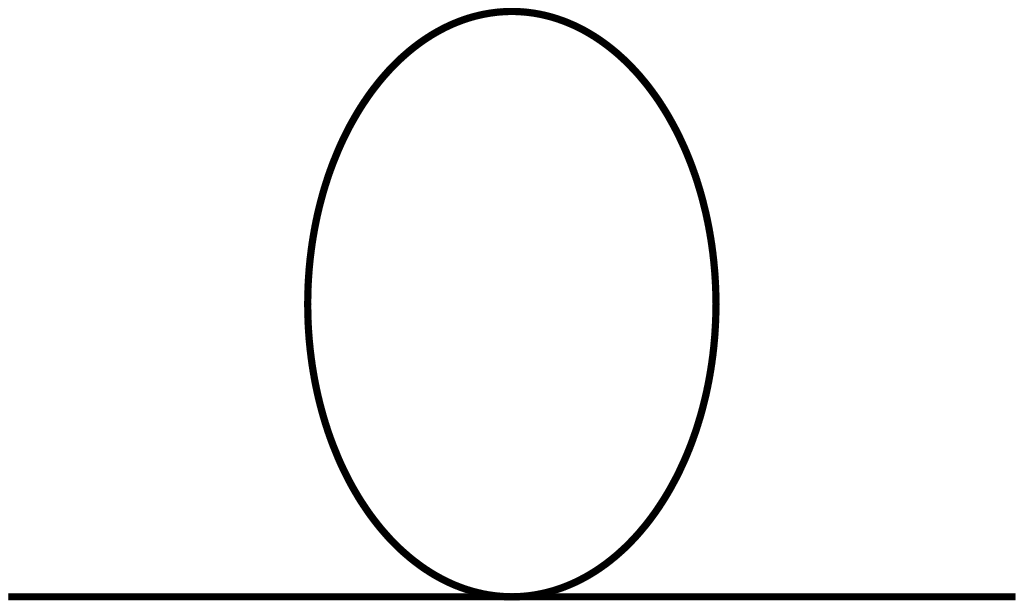}~~~~~~~~~~~~~~~~~~~~~~~
\includegraphics[width=4cm, clip]{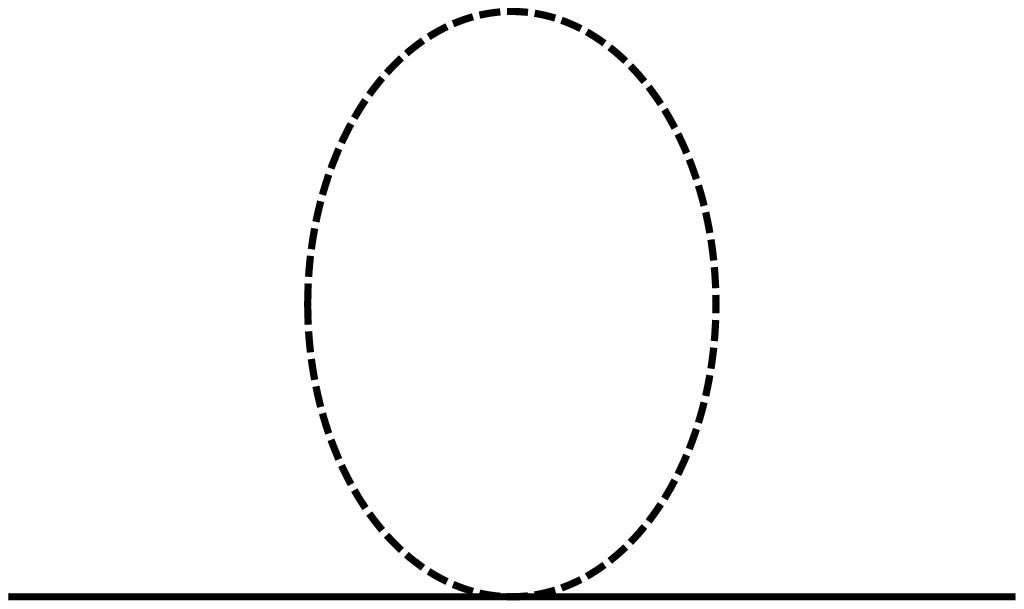}
\end{center}
\caption{One-loop diagrams contributing to the first order correction of $\alpha_1$. The solid lines represent the $\psi$ fields, and the dashed lines represent the $\phi$ fields. The external lines are slow modes, and the internal lines are fast modes.}
\label{aaaa1}
\end{figure}

 Two terms in the action (\ref{actioncoup}) contribute to the first-order corrections of $\alpha_1$. The coupling $ \psi_s^2  \psi_f^2$ leads to the correction
  \begin{equation}
 \delta^{(1)}S[\alpha_1]=3\beta_1\sum_{ijkl}F_{ijkl}\int_f \frac{d^{d+1}\mathbf q'}{(2\pi)^{d+1}} \langle\psi_f^i(\mathbf q') \psi_f^j(-\mathbf q') \rangle  \int_s \frac{d^{d+1}\mathbf q}{(2\pi)^{d+1}} \psi_s^k(\mathbf q) \psi_s^l(-\mathbf q).
 \end{equation}
 Using the identity,
 \begin{equation}
\sum_{i}F_{iikl}=\frac{n_1+2}{3}\delta_{kl},
\end{equation} 
 one obtains
 \begin{equation}
 \delta^{(1)}S[\alpha_1]=(n_1+2)\beta_1\frac{\Omega_d}{2\pi} \int d\omega \int_{b^{-1}}^1dkk^{d-1}   \frac{1}{-2\alpha_1+k^2+\gamma_1\frac{|\omega|}{k^{z_1-2}}} \int_s \frac{d^{d+1}\mathbf q}{(2\pi)^{d+1}}\boldsymbol \psi_s(\mathbf q)\cdot \boldsymbol \psi_s(-\mathbf q).
 \end{equation}
Assuming that the $\psi$ field is near its critical point, thus $\alpha_1$ is a small parameter, the Green's function can be expanded in terms of $-2\alpha_1$. The correction term can be written as
 \begin{equation}
 \delta^{(1)}S[\alpha_1]=(n_1+2)\beta_1\frac{\Omega_d}{2\pi} (I_1+2\alpha_1I_2) \int_s \frac{d^{d+1}\mathbf q}{(2\pi)^{d+1}}\boldsymbol \psi_s(\mathbf q)\cdot \boldsymbol \psi_s(-\mathbf q),
 \end{equation}
where we have defined the series of functions
\begin{equation}
I_n= \int  d\omega \int_{b^{-1}}^1dkk^{d-1}   \frac{1}{\left(k^2+\gamma_1\frac{|\omega|}{k^{z_1-2}}\right)^n}. 
\end{equation}
Let us first calculate $I_1$. The frequency integral requires a cutoff. From dimensional analysis, we choose to integrate over the region $-1<\gamma_1\omega<1$, and obtain the result
\begin{equation}
I_1= \frac{2}{\gamma_1} \int_{b^{-1}}^1dkk^{d+z_1-3} \ln  \left( \frac{1+k^{z_1}}{k^{z_1}}\right).
\label{caliii} 
\end{equation}
To proceed further, we are required to specify the dimension and dynamical exponent. Consider $d=3, z_1=2$, where one has 
\begin{equation}
I_1=\frac{2}{3\gamma_1}\left[ \ln2-b^{-3}\ln  \left( \frac{1+b^{-2}}{b^{-2}}\right)+2(1-b^{-1})-2\arctan 1+ 2\arctan b^{-1}  \right].
\end{equation}
Expanded to first order in $(1-b^{-1})$, it is simply
\begin{equation}
I_1=\frac{2\ln 2}{\gamma_1}(1-b^{-1}).
\label{iii}
\end{equation}
For $d=2, z_1=2$, we obtain
\begin{equation}
I_1=\frac{1}{\gamma_1}\left[ 2\ln2-(1+b^{-2})\ln (1+b^{-2})+b^{-2}\ln b^{-2}  \right],
\end{equation}
which leads to the same result (\ref{iii}) when expanded to first order in $(1-b^{-1})$. This result can also be obtained more crudely by setting $k=1$ in the integrand of (\ref{caliii}). $I_2$ can be calculated similarly, with the result
\begin{equation}
I_2=\frac{2}{\gamma_1}(1-b^{-1}).
\end{equation}
So the one-loop correction to $\alpha_1$ coming from the coupling $\psi_s^2 \psi_f^2$ is 
 \begin{equation}
 \delta^{(1)}S[\alpha_1]=(n_1+2)\beta_1\frac{\Omega_d}{\pi\gamma_1} (1-b^{-1})(\ln2+2\alpha_1) \int_s \frac{d^{d+1}\mathbf q}{(2\pi)^{d+1}}\boldsymbol \psi_s(\mathbf q)\cdot \boldsymbol \psi_s(-\mathbf q),
 \end{equation}

We next calculate contributions from the coupling $  \psi_s^2  \phi_f^2$, which takes the form
 \begin{equation}
 \delta^{(2)}S[\alpha_1]=g\sum_{ijkl}F'_{ijkl}\int_f \frac{d^{d+1}\mathbf q'}{(2\pi)^{d+1}} \langle\phi_f^i(\mathbf q') \phi_f^j(-\mathbf q') \rangle  \int_s \frac{d^{d+1}\mathbf q}{(2\pi)^{d+1}} \psi_s^k(\mathbf q) \psi_s^l(-\mathbf q),
 \end{equation}
with $F'_{ijkl}=\delta_{ij}\delta_{kl}$. So we have simply the identity
\begin{equation}
\sum_{i}F'_{iikl}=n_2\delta_{kl},
\end{equation}
which gives 
 \begin{equation}
 \delta^{(2)}S[\alpha_1]=n_2 g\frac{\Omega_d}{2\pi} \int d\omega \int_{b^{-1}}^1dkk^{d-1}   \frac{1}{-2\alpha_2+k^2+\omega^2} \int_s \frac{d^{d+1}\mathbf q}{(2\pi)^{d+1}}\boldsymbol \psi_s(\mathbf q)\cdot \boldsymbol \psi_s(-\mathbf q).
 \end{equation}
 Defining the new set of functions
 \begin{equation}
I'_n= \int  d\omega \int_{b^{-1}}^1dkk^{d-1}   \frac{1}{\left(k^2+\omega^2\right)^n}, 
\end{equation}
one obtains  
 \begin{equation}
 \delta^{(2)}S[\alpha_1]=n_2 g\frac{\Omega_d}{2\pi}  (I'_1+2\alpha_2 I'_2)   \int_s \frac{d^{d+1}\mathbf q}{(2\pi)^{d+1}}\boldsymbol \psi_s(\mathbf q)\cdot \boldsymbol \psi_s(-\mathbf q).
 \end{equation}
 Here we integrate over frequencies in the range $-\infty<\omega<\infty$, and get
 \begin{equation}
I'_1=\pi\int^1_{b^{-1}}dk k^{d-2},
\end{equation}
 which is, to first order in $(1-b^{-1})$,
\begin{equation}
I'_1=\pi(1-b^{-1}). 
\end{equation}
Similarly for $I'_2$ we have 
\begin{equation}
I'_2=\frac{\pi}{2}\int^1_{b^{-1}}dk k^{d-4},
\end{equation}
thus
\begin{equation}
I'_2=\frac{\pi}{2}(1-b^{-1}).
\end{equation}
Near $d=3$, one has $\Omega_d/4\simeq\Omega_{d+1}$. Grouping all these together, we obtain the second term in the correction to $\alpha_1$ as
 \begin{equation}
 \delta^{(2)}S[\alpha_1]=n_2 g \Omega_{d+1} (1-b^{-1}) (2+2\alpha_2)   \int_s \frac{d^{d+1}\mathbf q}{(2\pi)^{d+1}}\boldsymbol \psi_s(\mathbf q)\cdot \boldsymbol \psi_s(-\mathbf q).
 \end{equation}

 \subsection{First Order Corrections to $\alpha_2$}
 
 \begin{figure}[ht]
\begin{center}
\includegraphics[width=4cm, clip]{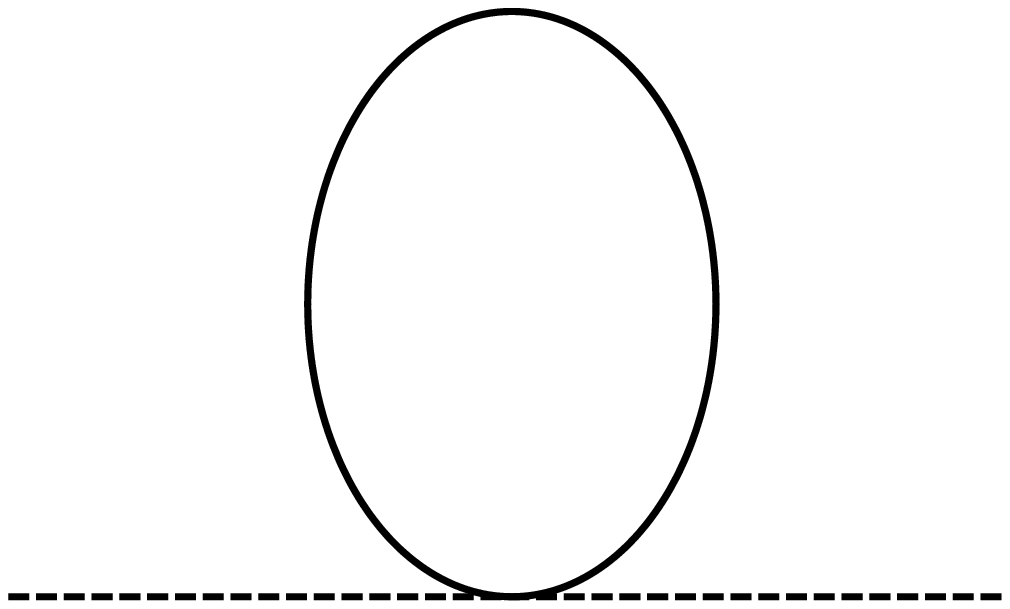}~~~~~~~~~~~~~~~~~~~~~~~
\includegraphics[width=4cm, clip]{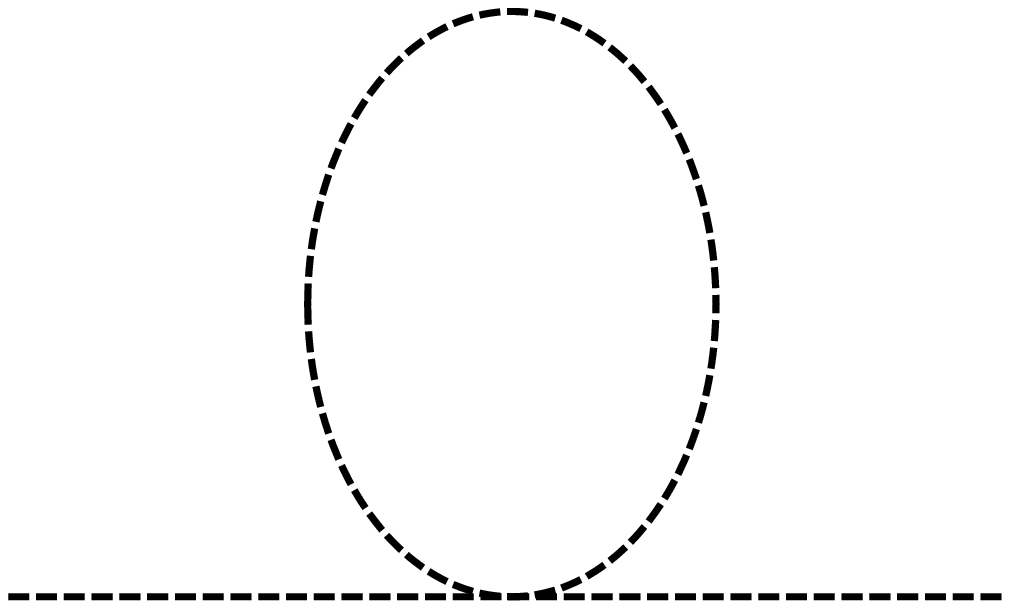}
\end{center}
\caption{One-loop diagrams contributing to the first order correction of $\alpha_2$. The solid lines represent the $\psi$ fields, and the dashed lines represent the $\phi$ fields. The external lines are slow modes, and the internal lines are fast modes.}
\label{aaaa2}
\end{figure}

 The calculation of the first order corrections to $\alpha_2$ is quite similar to that of $\alpha_1$. There are again two terms contributing. The coupling $\psi_f^2  \phi_s^2$
 gives rise to a term of the form
  \begin{equation}
 \delta^{(1)}S[\alpha_2]=g\sum_{ijkl}F'_{ijkl}\int_f \frac{d^{d+1}\mathbf q'}{(2\pi)^{d+1}} \langle\psi_f^i(\mathbf q') \psi_f^j(-\mathbf q') \rangle  \int_s \frac{d^{d+1}\mathbf q}{(2\pi)^{d+1}} \phi_s^k(\mathbf q) \phi_s^l(-\mathbf q),
 \end{equation}
Summing over the field indices, 
  \begin{equation}
\sum_{i}F'_{iikl}\phi_s^k \phi_s^l=n_1 |\boldsymbol \phi_s|^2,
 \end{equation}
 we obtain
  \begin{equation}
 \delta^{(1)}S[\alpha_2]=n_1g\frac{\Omega_d}{2\pi} \int d\omega \int_{b^{-1}}^1dkk^{d-1}   \frac{1}{-2\alpha_1+k^2+\gamma_1\frac{|\omega|}{k^{z_1-2}}} \int_s \frac{d^{d+1}\mathbf q}{(2\pi)^{d+1}}\boldsymbol \phi_s(\mathbf q)\cdot \boldsymbol \phi_s(-\mathbf q),
 \end{equation}
 which can be expanded as
  \begin{equation}
 \delta^{(1)}S[\alpha_2]=n_1g\frac{\Omega_d}{2\pi} (I_1+2\alpha_1I_2) \int_s \frac{d^{d+1}\mathbf q}{(2\pi)^{d+1}}\boldsymbol \phi_s(\mathbf q)\cdot \boldsymbol \phi_s(-\mathbf q).
 \end{equation}
 The result is 
  \begin{equation}
 \delta^{(1)}S[\alpha_2]=n_1g \frac{\Omega_d}{\pi\gamma_1} (1-b^{-1})(\ln2+2\alpha_1) 
   \int_s \frac{d^{d+1}\mathbf q}{(2\pi)^{d+1}}\boldsymbol \phi_s(\mathbf q)\cdot \boldsymbol \phi_s(-\mathbf q).
 \end{equation}
 
The other term comes from the coupling $\phi_f^2  \phi_s^2$. It has the form
 \begin{equation}
 \delta^{(2)}S[\alpha_2]=3\beta_2\sum_{ijkl}F_{ijkl}\int_f \frac{d^{d+1}\mathbf q'}{(2\pi)^{d+1}} \langle\phi_f^i(\mathbf q') \phi_f^j(-\mathbf q') \rangle  \int_s \frac{d^{d+1}\mathbf q}{(2\pi)^{d+1}} \phi_s^k(\mathbf q) \phi_s^l(-\mathbf q).
 \end{equation}
 We first sum over the field indices,
   \begin{equation}
\sum_{i}F_{iikl}\phi_s^k \phi_s^l=\frac{n_2+2}{3} |\boldsymbol \phi_s|^2,
 \end{equation}
 resulting in
  \begin{equation}
 \delta^{(2)}S[\alpha_2]=(n_2+2) \beta_2\frac{\Omega_d}{2\pi} \int d\omega \int_{b^{-1}}^1dkk^{d-1}   \frac{1}{-2\alpha_2+k^2+\omega^2} \int_s \frac{d^{d+1}\mathbf q}{(2\pi)^{d+1}}\boldsymbol \phi_s(\mathbf q)\cdot \boldsymbol \phi_s(-\mathbf q).
 \end{equation}
 Expanding to first order in $\alpha_2$, one has 
   \begin{equation}
 \delta^{(2)}S[\alpha_2]=(n_2+2) \beta_2\frac{\Omega_d}{2\pi}(I'_1+2\alpha_2 I'_2)
 \int_s \frac{d^{d+1}\mathbf q}{(2\pi)^{d+1}}\boldsymbol \phi_s(\mathbf q)\cdot \boldsymbol \phi_s(-\mathbf q),
 \end{equation}
 and the final result is
    \begin{equation}
 \delta^{(2)}S[\alpha_2]=(n_2+2) \beta_2 \Omega_{d+1} (1-b^{-1}) (2+2\alpha_2) 
 \int_s \frac{d^{d+1}\mathbf q}{(2\pi)^{d+1}}\boldsymbol \phi_s(\mathbf q)\cdot \boldsymbol \phi_s(-\mathbf q),
 \end{equation}

 \subsection{First Order Corrections to $\beta_1$}
 
 \begin{figure}[ht]
\begin{center}
\includegraphics[width=4cm, clip]{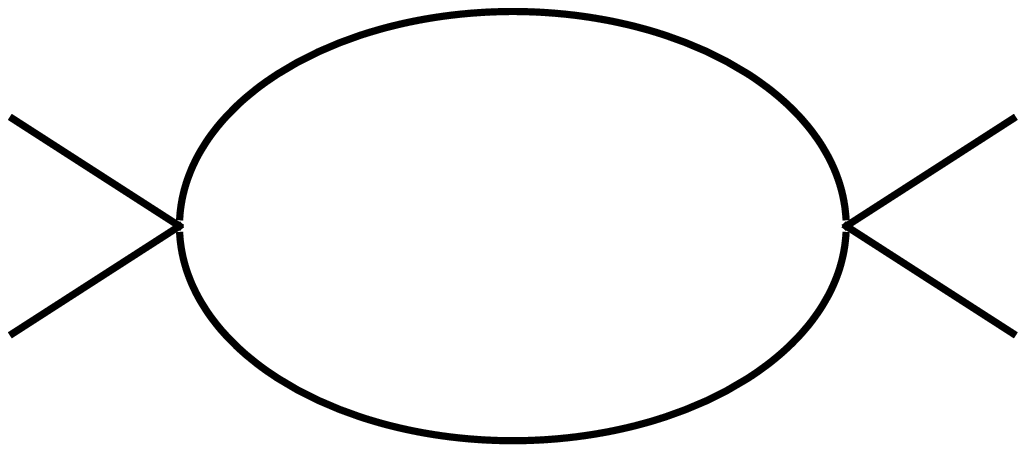}~~~~~~~~~~~~~~~~~~~~~~~
\includegraphics[width=4cm, clip]{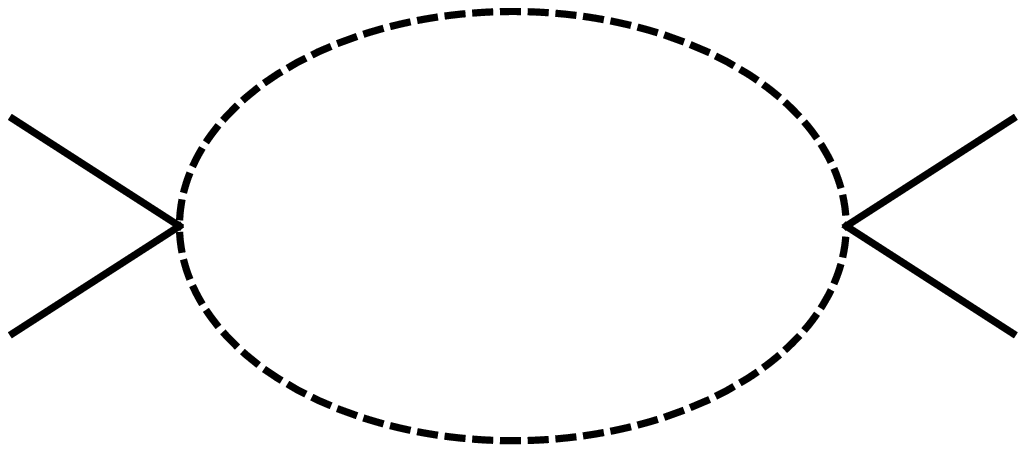}
\end{center}
\caption{One-loop diagrams contributing to the first order correction of $\beta_1$. The solid lines represent the $\psi$ fields, and the dashed lines represent the $\phi$ fields. The external lines are slow modes, and the internal lines are fast modes.}
\label{bbbb1}
\end{figure}
 
 The first order correction to $\beta_1$ comes from two one-loop diagrams, one with two internal $\psi_f$ lines, the other with two $\phi_f$ lines. The dependence of the internal lines on the external momenta and frequencies can be ignored here, since they are of higher order. 
 
 The first term with $\psi_f$ internal lines is of the form 
   \begin{equation}
    \begin{split}
\\ \delta^{(1)}S[\beta_1]=&-(3\beta_1)^2\sum_{k_1k_2l_1l_2}\sum_{i_1i_2j_1j_2}F_{i_1j_1k_1l_1}F_{i_2j_2k_2l_2}\int_f \frac{d^{d+1}\mathbf q'}{(2\pi)^{d+1}} \langle\psi_f^{i_1}(\mathbf q') \psi_f^{i_2}(-\mathbf q') \rangle \langle\psi_f^{j_1}(\mathbf q') \psi_f^{j_2}(-\mathbf q') \rangle \\&\times\int d^d{\mathbf r}d\tau \psi_s^{k_1} \psi_s^{k_2} \psi_s^{l_1} \psi_s^{l_2}
+~2 ~{\rm permutations}.
 \end{split}
 \end{equation}
Using the identity,
 \begin{equation}
\sum_{ij}F_{ijk_1l_1}F_{ijk_2l_2}=\frac{1}{9}\left[(n_1+4)\delta_{k_1l_1}\delta_{k_2l_2}+2\delta_{k_1k_2}\delta_{l_1l_2}+2\delta_{k_1l_2}\delta_{k_2l_1} \right],
\end{equation} 
combined with the 2 other permutations of the external lines, the part containing the field component indices can be simplified as
 \begin{equation}
\sum_{k_1k_2l_1l_2}\sum_{ij}F_{ijk_1l_1}F_{ijk_2l_2} \psi_s^{k_1} \psi_s^{k_2} \psi_s^{l_1} \psi_s^{l_2} +~2 ~{\rm permutations}= \frac{n_1+8}{9}|\boldsymbol \psi_s|^4.
\end{equation} 
Thus the first correction to $\beta_1$ reads
  \begin{equation}
 \delta^{(1)}S[\beta_1]=-(n_1+8)\beta_1^2 \frac{\Omega_d}{2\pi} \int d\omega \int_{b^{-1}}^1dkk^{d-1}   \frac{1}{(-2\alpha_1+k^2+\gamma_1\frac{|\omega|}{k^{z_1-2}})^2}      \int d^d{\mathbf r}d\tau |\boldsymbol \psi_s|^4,
 \end{equation}
which is, to leading order of $\alpha_1$,
  \begin{equation}
 \delta^{(1)}S[\beta_1]=-(n_1+8)\beta_1^2 \frac{\Omega_d}{2\pi} I_2   \int d^d{\mathbf r}d\tau |\boldsymbol \psi_s|^4.
 \end{equation}
 Substituting the explicit expression for $I_2$, we get the result
   \begin{equation}
 \delta^{(1)}S[\beta_1]=-(n_1+8)\beta_1^2 \frac{\Omega_d}{\pi\gamma_1} (1-b^{-1})   \int d^d{\mathbf r}d\tau |\boldsymbol \psi_s|^4.
 \end{equation}

The second term has two $\phi_f$ internal lines, and takes the form
  \begin{equation}
    \begin{split}
\\ \delta^{(2)}S[\beta_1]=&-g^2   \sum_{i_1i_2j_1j_2}\sum_{k_1k_2l_1l_2}    F'_{i_1j_1k_1l_1}F'_{i_2j_2k_2l_2}\int_f \frac{d^{d+1}\mathbf q'}{(2\pi)^{d+1}} \langle\phi_f^{i_1}(\mathbf q') \phi_f^{i_2}(-\mathbf q') \rangle \langle\phi_f^{j_1}(\mathbf q') \phi_f^{j_2}(-\mathbf q') \rangle\\&\times \int d^d{\mathbf r}d\tau \psi_s^{k_1} \psi_s^{k_2} \psi_s^{l_1} \psi_s^{l_2}
+~2 ~{\rm permutations}.
 \end{split}
 \end{equation}
The part with the field component indices gives
 \begin{equation}
\sum_{k_1k_2l_1l_2}\sum_{ij}F'_{ijk_1l_1}F'_{ijk_2l_2} \psi_s^{k_1} \psi_s^{k_2} \psi_s^{l_1} \psi_s^{l_2} +~2 ~{\rm permutations}=n_2 |\boldsymbol \psi_s|^4,
\end{equation} 
which further leads to the result
  \begin{equation}
 \delta^{(2)}S[\beta_1]=-n_2g^2 \frac{\Omega_d}{2\pi} \int d\omega \int_{b^{-1}}^1dkk^{d-1}   \frac{1}{(-2\alpha_2+k^2+\omega^2)^2}      \int d^d{\mathbf r}d\tau |\boldsymbol \psi_s|^4.
 \end{equation}
To leading order in $\alpha_2$, it is 
   \begin{equation}
 \delta^{(2)}S[\beta_1]=-n_2g^2 \frac{\Omega_d}{2\pi} I'_2   \int d^d{\mathbf r}d\tau |\boldsymbol \psi_s|^4,
 \end{equation}
or more explicitly,
   \begin{equation}
 \delta^{(2)}S[\beta_1]=-n_2g^2 \Omega_{d+1}(1-b^{-1})   \int d^d{\mathbf r}d\tau |\boldsymbol \psi_s|^4.
 \end{equation}

 \subsection{First Order Corrections to $\beta_2$}
 
  \begin{figure}[ht]
\begin{center}
\includegraphics[width=4cm, clip]{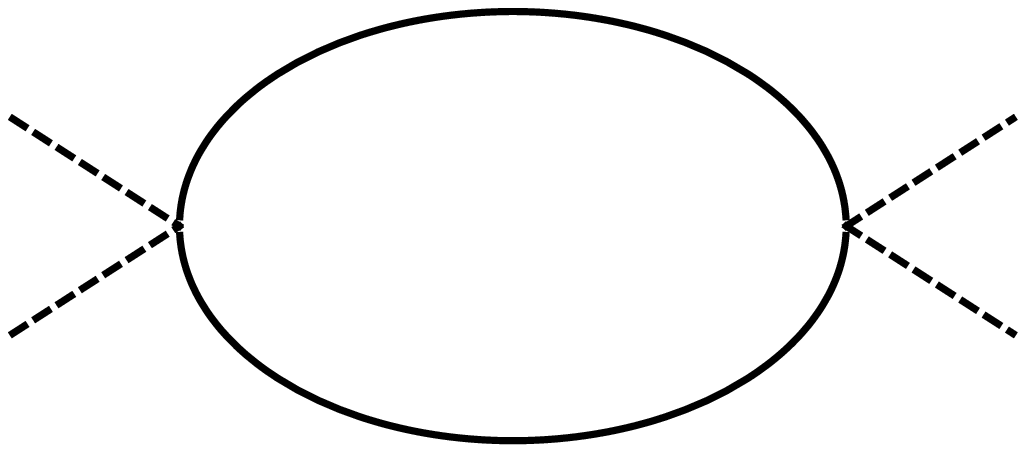}~~~~~~~~~~~~~~~~~~~~~~~
\includegraphics[width=4cm, clip]{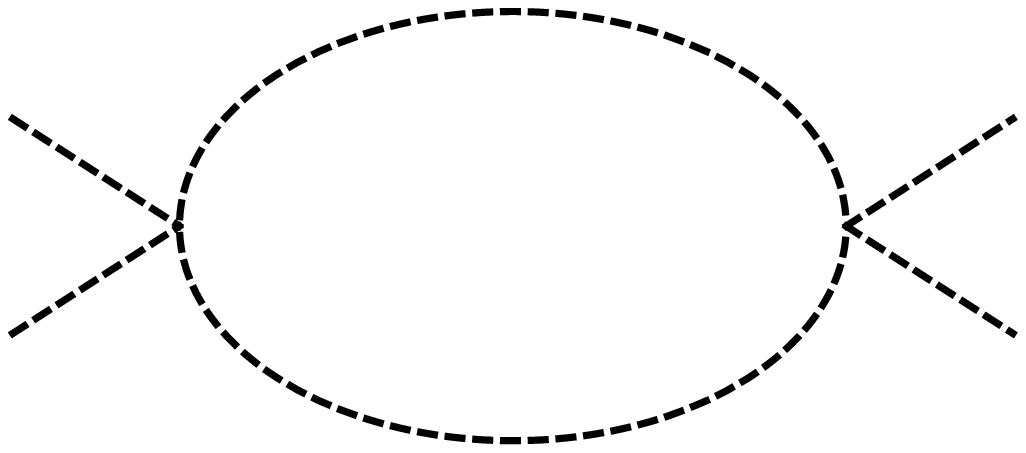}
\end{center}
\caption{One-loop diagrams contributing to the first order correction of $\beta_2$. The solid lines represent the $\psi$ fields, and the dashed lines represent the $\phi$ fields. The external lines are slow modes, and the internal lines are fast modes.}
\label{bbbb2}
\end{figure}

The first order correction to the $\beta_2$ term also comes from two diagrams. The first one has two $\psi_f$ internal lines, and is of the form
 \begin{equation}
    \begin{split}
\\ \delta^{(1)}S[\beta_2]=&-g^2\sum_{i_1i_2j_1j_2}\sum_{k_1k_2l_1l_2}F'_{i_1j_1k_1l_1}F'_{i_2j_2k_2l_2}\int_f \frac{d^{d+1}\mathbf q'}{(2\pi)^{d+1}} \langle\psi_f^{i_1}(\mathbf q') \psi_f^{i_2}(-\mathbf q') \rangle \langle\psi_f^{j_1}(\mathbf q') \psi_f^{j_2}(-\mathbf q') \rangle\\&\times \int d^d{\mathbf r}d\tau \phi_s^{k_1} \phi_s^{k_2} \phi_s^{l_1} \phi_s^{l_2}
+~2 ~{\rm permutations}.
 \end{split}
 \end{equation}
Summing over different field components, where one has
 \begin{equation}
\sum_{k_1k_2l_1l_2}\sum_{ij}F'_{ijk_1l_1}F'_{ijk_2l_2} \phi_s^{k_1} \phi_s^{k_2} \phi_s^{l_1} \phi_s^{l_2} +~2 ~{\rm permutations}=n_1 |\boldsymbol \phi_s|^4,
\end{equation} 
the first correction to the $\beta_2$ term is 
  \begin{equation}
 \delta^{(1)}S[\beta_2]=-n_1g^2 \frac{\Omega_d}{2\pi} \int d\omega \int_{b^{-1}}^1dkk^{d-1}   \frac{1}{(-2\alpha_1+k^2+\gamma_1\frac{|\omega|}{k^{z_1-2}})^2}      \int d^d{\mathbf r}d\tau |\boldsymbol \phi_s|^4.
 \end{equation}
To first order in $\alpha_1$, it is simply 
  \begin{equation}
 \delta^{(1)}S[\beta_2]=-n_1g^2 \frac{\Omega_d}{2\pi}I_2     \int d^d{\mathbf r}d\tau |\boldsymbol \phi_s|^4,
 \end{equation}
which can be written as
  \begin{equation}
 \delta^{(1)}S[\beta_2]=-n_1g^2 \frac{\Omega_d}{\pi\gamma_1} (1-b^{-1})    \int d^d{\mathbf r}d\tau |\boldsymbol \phi_s|^4.
 \end{equation}

The second diagram contains two $\phi_f$ internal lines, thus the correction reads
 \begin{equation}
 \begin{split}
\\ \delta^{(2)}S[\beta_2]=&-(3\beta_2)^2\sum_{k_1k_2l_1l_2}\sum_{i_1i_2j_1j_2}F_{i_1j_1k_1l_1}F_{i_2j_2k_2l_2}\int_f \frac{d^{d+1}\mathbf q'}{(2\pi)^{d+1}} \langle\phi_f^{i_1}(\mathbf q') \phi_f^{i_2}(-\mathbf q') \rangle \langle\phi_f^{j_1}(\mathbf q') \phi_f^{j_2}(-\mathbf q') \rangle \\&\times\int d^d{\mathbf r}d\tau \phi_s^{k_1} \phi_s^{k_2} \phi_s^{l_1} \phi_s^{l_2}
+~2 ~{\rm permutations}.
 \end{split}
 \end{equation}
 The summation over the field indices gives 
  \begin{equation}
\sum_{k_1k_2l_1l_2}\sum_{ij}F_{ijk_1l_1}F_{ijk_2l_2} \phi_s^{k_1} \phi_s^{k_2} \phi_s^{l_1} \phi_s^{l_2} +~2 ~{\rm permutations}= \frac{n_2+8}{9}|\boldsymbol \phi_s|^4.
\end{equation} 
Thus the second correction to $\beta_2$ reads
  \begin{equation}
 \delta^{(2)}S[\beta_2]=-(n_2+8)\beta_2^2 \frac{\Omega_d}{2\pi} \int d\omega \int_{b^{-1}}^1dkk^{d-1}   \frac{1}{(-2\alpha_2+k^2+\omega^2)^2}      \int d^d{\mathbf r}d\tau |\boldsymbol \phi_s|^4.
 \end{equation}
 When the $\phi$ field is near its critical point, the above expression can be simplified to be
   \begin{equation}
 \delta^{(2)}S[\beta_2]=-(n_2+8)\beta_2^2 \frac{\Omega_d}{2\pi} I'_2   \int d^d{\mathbf r}d\tau |\boldsymbol \phi_s|^4,
 \end{equation}
 which is 
    \begin{equation}
 \delta^{(2)}S[\beta_2]=-(n_2+8)\beta_2^2 \Omega_{d+1}(1-b^{-1})  \int d^d{\mathbf r}d\tau |\boldsymbol \phi_s|^4.
 \end{equation}
 
 \subsection{First Order Corrections to $g$}
 
  \begin{figure}[ht]
\begin{center}
\includegraphics[width=4cm, clip]{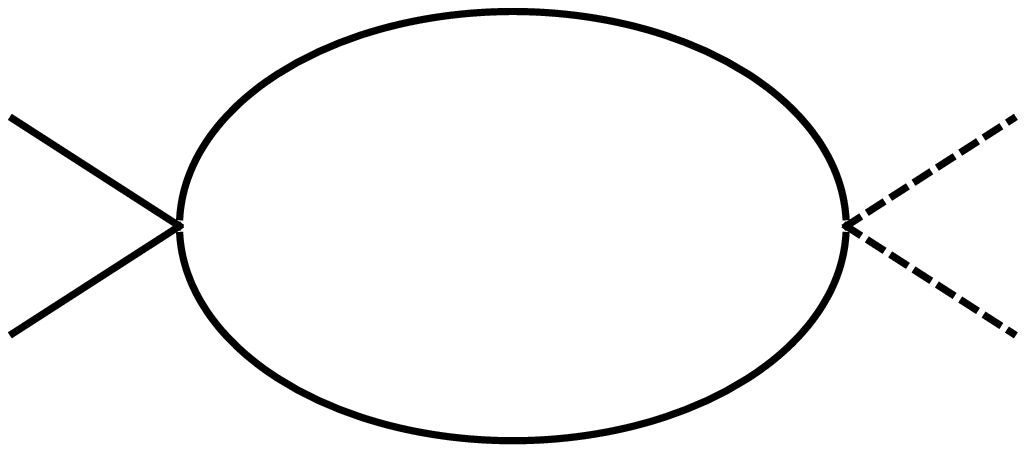} ~~~~~~~~~~~
\includegraphics[width=4cm, clip]{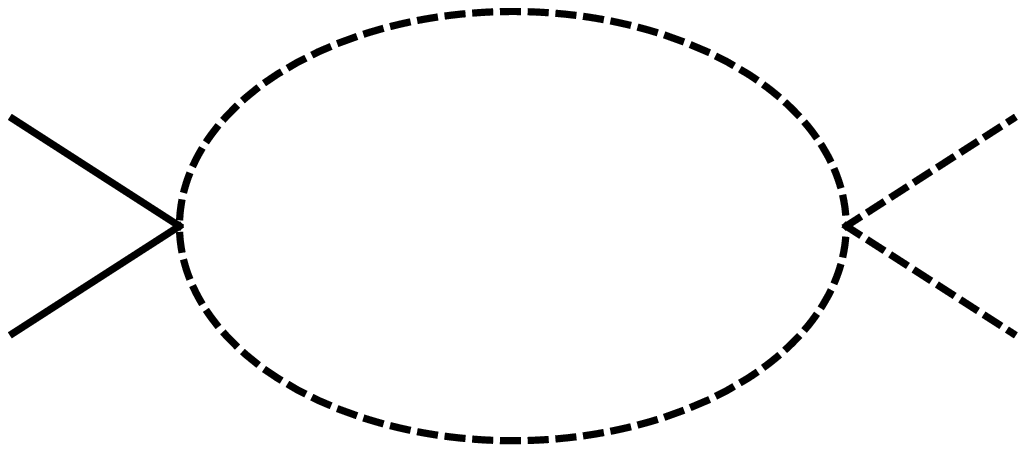} ~~~~~~~~~~~
\includegraphics[width=4cm, clip]{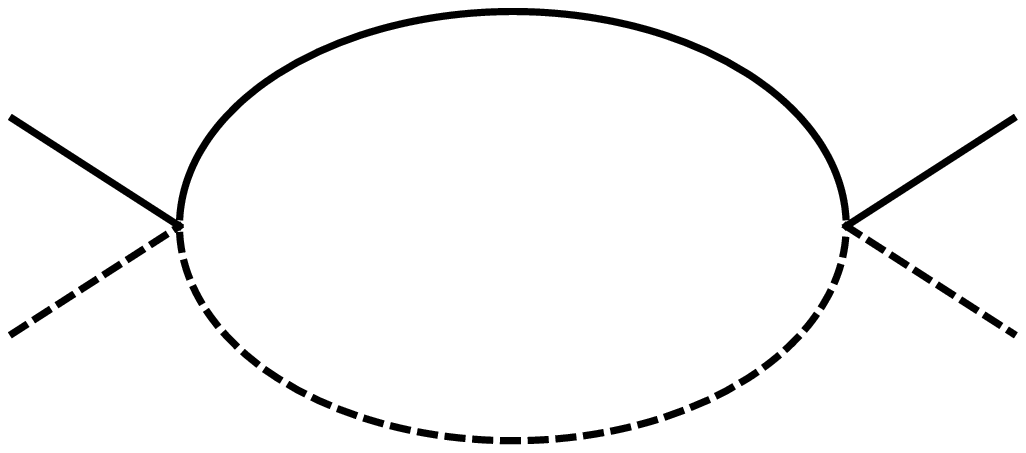}
\end{center}
\caption{One-loop diagrams contributing to the first order correction of $g$. The solid lines represent the $\psi$ fields, and the dashed lines represent the $\phi$ fields. The external lines are slow modes, and the internal lines are fast modes.}
\label{gggg}
\end{figure}

 There are three diagrams contributing to the first order corrections of the coupling $g$ between the squares of the two fields. The first diagram has two $\psi_f$ fields as internal lines. This term takes the form
   \begin{equation}
    \begin{split}
\delta^{(1)}S[g]=&-\frac{1}{2}\times 2\times 2(3\beta_1)g\sum_{k_1k_2l_1l_2}\sum_{i_1i_2j_1j_2}F_{i_1j_1k_1l_1}F'_{i_2j_2k_2l_2}\int_f \frac{d^{d+1}\mathbf q'}{(2\pi)^{d+1}} \langle\psi_f^{i_1}(\mathbf q') \psi_f^{i_2}(-\mathbf q') \rangle \langle\psi_f^{j_1}(\mathbf q') \psi_f^{j_2}(-\mathbf q') \rangle\\&\times \int d^d{\mathbf r}d\tau \psi_s^{k_1} \psi_s^{l_1} \phi_s^{k_2} \phi_s^{l_2}.
 \end{split}
 \end{equation} 
 The $1/2$ comes from $(1/2)S_c^2$, and the two $2$ factors come from the expansion in $S_c^2$ and the number of contractions in $\langle  \psi_f\psi_f(x)\psi_f\psi_f(y) \rangle$. 
 We first sum over the field indices, 
   \begin{equation}
\sum_{k_1k_2l_1l_2}\sum_{ij}F_{ijk_1l_1}F'_{ijk_2l_2} \psi_s^{k_1} \psi_s^{l_1} \phi_s^{k_2} \phi_s^{l_2}=\frac{n_1+2}{3} |\boldsymbol \psi_s|^2  |\boldsymbol \phi_s|^2,
 \end{equation} 
and then substitute the Green's functions,
  \begin{equation}
\delta^{(1)}S[g]=-2\beta_1g (n_1+2) \frac{\Omega_d}{2\pi} \int d\omega \int_{b^{-1}}^1dkk^{d-1}   \frac{1}{(-2\alpha_1+k^2+\gamma_1\frac{|\omega|}{k^{z_1-2}})^2}   
 \int d^d{\mathbf r}d\tau |\boldsymbol \psi_s|^2  |\boldsymbol \phi_s|^2.
 \end{equation} 
 Keeping only the leading order term, 
   \begin{equation}
\delta^{(1)}S[g]=-2\beta_1g (n_1+2) \frac{\Omega_d}{2\pi} I_2
 \int d^d{\mathbf r}d\tau |\boldsymbol \psi_s|^2  |\boldsymbol \phi_s|^2,
 \end{equation}
 one arrives at the result,
    \begin{equation}
\delta^{(1)}S[g]=-2\beta_1g (n_1+2) \frac{\Omega_d}{\pi\gamma_1} (1-b^{-1})
 \int d^d{\mathbf r}d\tau |\boldsymbol \psi_s|^2  |\boldsymbol \phi_s|^2.
 \end{equation}
 
 The internal lines of the second diagram are two $\phi_f$ fields. The corresponding correction term  is now
  \begin{equation}
    \begin{split}
\delta^{(2)}S[g]=&-\frac{1}{2}\times 2\times 2(3\beta_2)g\sum_{k_1k_2l_1l_2}\sum_{i_1i_2j_1j_2}F_{i_1j_1k_1l_1}F'_{i_2j_2k_2l_2}\int_f \frac{d^{d+1}\mathbf q'}{(2\pi)^{d+1}} \langle\phi_f^{i_1}(\mathbf q') \phi_f^{i_2}(-\mathbf q') \rangle \langle\phi_f^{j_1}(\mathbf q') \phi_f^{j_2}(-\mathbf q') \rangle\\&\times \int d^d{\mathbf r}d\tau \phi_s^{k_1} \phi_s^{l_1} \psi_s^{k_2} \psi_s^{l_2}.
 \end{split}
 \end{equation} 
 The summation over field indices is similar to the first term,  
  \begin{equation}
\sum_{k_1k_2l_1l_2}\sum_{ij}F_{ijk_1l_1}F'_{ijk_2l_2} \phi_s^{k_1} \phi_s^{l_1} \psi_s^{k_2} \psi_s^{l_2}=\frac{n_2+2}{3} |\boldsymbol \psi_s|^2  |\boldsymbol \phi_s|^2.
 \end{equation} 
 Thus the correction to the action is also similar,
   \begin{equation}
\delta^{(2)}S[g]=-2\beta_2g (n_2+2) \frac{\Omega_d}{2\pi} \int d\omega \int_{b^{-1}}^1dkk^{d-1}   \frac{1}{(-2\alpha_2+k^2+\omega^2)^2}   
 \int d^d{\mathbf r}d\tau |\boldsymbol \psi_s|^2  |\boldsymbol \phi_s|^2,
 \end{equation} 
 which is, to leading order in $\alpha_2$,
    \begin{equation}
\delta^{(2)}S[g]=-2\beta_2g (n_2+2) \frac{\Omega_d}{2\pi} I'_2 \int d^d{\mathbf r}d\tau |\boldsymbol \psi_s|^2  |\boldsymbol \phi_s|^2,
 \end{equation} 
 or
  \begin{equation}
\delta^{(2)}S[g]=-2\beta_2g (n_2+2) \Omega_{d+1}(1-b^{-1}) \int d^d{\mathbf r}d\tau |\boldsymbol \psi_s|^2  |\boldsymbol \phi_s|^2.
 \end{equation}  
 
 The third diagram has one $\phi_f$ internal line, and one $\psi_f$ internal line. The correction takes the form 
   \begin{equation}
    \begin{split}
\delta^{(3)}S[g]=&-\frac{1}{2}(4g)^2\sum_{k_1k_2l_1l_2}\sum_{i_1i_2j_1j_2}F'_{i_1j_1k_1l_1}F'_{i_2j_2k_2l_2}\int_f \frac{d^{d+1}\mathbf q'}{(2\pi)^{d+1}} \langle\psi_f^{i_1}(\mathbf q') \psi_f^{i_2}(-\mathbf q') \rangle \langle\phi_f^{j_1}(\mathbf q') \phi_f^{j_2}(-\mathbf q') \rangle\\&\times \int d^d{\mathbf r}d\tau \psi_s^{k_1} \phi_s^{l_1} \psi_s^{k_2} \phi_s^{l_2}.
 \end{split}
 \end{equation} 
 With the summation
  \begin{equation}
\sum_{k_1k_2l_1l_2}\sum_{ij}F'_{ik_1jl_1}F'_{ik_2jl_2} \psi_s^{k_1} \phi_s^{l_1} \psi_s^{k_2} \phi_s^{l_2}= |\boldsymbol \psi_s|^2  |\boldsymbol \phi_s|^2,
 \end{equation} 
 we obtain
   \begin{equation}
\delta^{(3)}S[g]=-8g^2 \frac{\Omega_d}{2\pi} \int d\omega \int_{b^{-1}}^1dkk^{d-1}   \frac{1}{-2\alpha_2+k^2+\omega^2}   \frac{1}{-2\alpha_1+k^2+\gamma_1\frac{|\omega|}{k^{z_1-2}}} 
 \int d^d{\mathbf r}d\tau |\boldsymbol \psi_s|^2  |\boldsymbol \phi_s|^2.
 \end{equation} 
 Assuming both fields are near their critical points, the above equation is approximately 
 \begin{equation}
\delta^{(3)}S[g]=-8g^2 \frac{\Omega_d}{2\pi}I''
 \int d^d{\mathbf r}d\tau |\boldsymbol \psi_s|^2  |\boldsymbol \phi_s|^2,
 \end{equation}  
 with the new function $I''$ defined as 
    \begin{equation}
I''=\int _{-\infty}^{\infty} d\omega\int_{b^{-1}}^1dkk^{d-1}   \frac{1}{k^2+\omega^2}   \frac{1}{k^2+\gamma_1\frac{|\omega|}{k^{z_1-2}}} .
 \end{equation} 
 As mentioned before, here in our RG scheme, frequency is integrated over the whole real axes. In the two propagators, frequency scales differently with momentum. For the $\psi$ field, $\gamma_1\omega\sim k^{z_1}$; for the $\phi$ field, $\omega\sim k$. And a finite cut-off in frequency would lead to inconsistencies for such cases with miscellaneous dynamical exponents.
 Here the frequency integral gives 
     \begin{equation}
I''=\int_{b^{-1}}^1dkk^{d+z_1-3}\frac{1}{\gamma_1^2k^2+k^{2z_1}}\left(  -\gamma_1\ln\frac{k^{2z_1-2}}{\gamma_1^2} +\pi k^{z_1-1}  \right).
 \end{equation} 
To leading order in $(1-b^{-1})$, we have 
     \begin{equation}
I''=\frac{1}{1+\gamma_1^2}(1-b^{-1})\left( 2 \gamma_1\ln{\gamma_1} +\pi  \right).
 \end{equation} 
 This leads to the third term in the correction to the $g$ term
  \begin{equation}
\delta^{(3)}S[g]=-8g^2 \frac{\Omega_d}{2\pi}\frac{ 2 \gamma_1\ln{\gamma_1} +\pi  }{1+\gamma_1^2}(1-b^{-1})
 \int d^d{\mathbf r}d\tau |\boldsymbol \psi_s|^2  |\boldsymbol \phi_s|^2,
 \end{equation} 
 
 \subsection{Rescaling of the Parameters}
 Now combining all the above results for the corrections of the different parameters, and carrying out the rescaling 
  \begin{equation}
 \begin{split}
\\ {\mathbf r}\to & {\mathbf r}/b,
\\ \tau\to&\tau/b,
\\ \boldsymbol \psi\to&  b^{(d-1)/2}\boldsymbol \psi,
\\ \boldsymbol \phi\to& b^{(d-1)/2}\boldsymbol \phi,
\end{split}
\label{dim2}
\end{equation}
 we obtain the RG equations
  \begin{equation}
 \begin{split}
\\ \gamma_1\to & b^{z-1}\gamma_1,
\\ -\alpha_1\to&b^{2}\left[ -\alpha_1+(n_1+2)\beta_1\frac{\Omega_d}{\pi\gamma_1} (1-b^{-1})(\ln2+2\alpha_1) +n_2 g \Omega_{d+1} (1-b^{-1}) (2+2\alpha_2) \right],
\\ -\alpha_2\to&b^{2}\left[ -\alpha_2+ (n_2+2) \beta_2 \Omega_{d+1} (1-b^{-1}) (2+2\alpha_2)   +n_1g \frac{\Omega_d}{\pi\gamma_1} (1-b^{-1})(\ln2+2\alpha_1)  \right],
\\ \frac{\beta_1}{2} \to & b^\epsilon \left[  \frac{\beta_1}{2}- (n_1+8)\beta_1^2 \frac{\Omega_d}{\pi\gamma_1} (1-b^{-1}) -n_2g^2 \Omega_{d+1}(1-b^{-1})   \right],
\\ \frac{\beta_2}{2} \to & b^\epsilon \left[ \frac{\beta_2}{2}-n_1g^2 \frac{\Omega_d}{\pi\gamma_1} (1-b^{-1})    -(n_2+8)\beta_2^2 \Omega_{d+1}(1-b^{-1}) \right],
\\ g\to & b^\epsilon  \left[  g-2\beta_1g (n_1+2) \frac{\Omega_d}{\pi\gamma_1} (1-b^{-1})-2\beta_2g (n_2+2) \Omega_{d+1}(1-b^{-1})  -8g^2 \frac{\Omega_d}{2\pi}\frac{ 2 \gamma_1\ln{\gamma_1} +\pi  }{1+\gamma_1^2} (1-b^{-1})\right],
\end{split}
\end{equation} 
 the differential form of which has been presented in equations (\ref{rgg}, \ref{rgab}).

\bibliographystyle{apsrev}
\bibliography{strings,refs}

\end{document}